\documentclass[a4paper,prl,twocolumn,superscriptaddress]{revtex4}
\usepackage{graphicx}
\usepackage{amsmath}
\usepackage{amssymb}
\usepackage{float}
\usepackage[utf8]{inputenc}

\begin{document}

\title{Collapse of the Josephson emission in a carbon nanotube junction in the Kondo regime}
\author{D. Watfa}
\affiliation{Universit\'e Paris-Saclay, CNRS, Laboratoire de Physique des Solides, 91405, Orsay, France.}
\author{R. Delagrange}
\affiliation{Universit\'e Paris-Saclay, CNRS, Laboratoire de Physique des Solides, 91405, Orsay, France.}
\author{A. Kadlecov\'a}
\affiliation{Department of Condensed Matter Physics, Faculty of Mathematics and Physics, Charles University, Ke Karlovu 5,
CZ-121 16 Praha 2, Czech Republic}
\author{M. Ferrier}
\affiliation{Universit\'e Paris-Saclay, CNRS, Laboratoire de Physique des Solides, 91405, Orsay, France.}
\author{A. Kasumov}
\affiliation{Universit\'e Paris-Saclay, CNRS, Laboratoire de Physique des Solides, 91405, Orsay, France.}
\author{H. Bouchiat}
\affiliation{Universit\'e Paris-Saclay, CNRS, Laboratoire de Physique des Solides, 91405, Orsay, France.}
\author{R. Deblock}
\affiliation{Universit\'e Paris-Saclay, CNRS, Laboratoire de Physique des Solides, 91405, Orsay, France.}

\begin{abstract}
We probe the high frequency emission of a carbon nanotube based Josephson junction and compare it to its DC Josephson current. The AC emission is probed by coupling the carbon nanotube to an on-chip detector (a Superconductor-Insulator-Superconductor junction), via a coplanar waveguide resonator. The measurement of the photo-assisted current of the detector gives direct access to the signal emitted by the carbon nanotube. We focus on the gate regions that exhibit Kondo features in the normal state and demonstrate that when the DC supercurrent is enhanced by the Kondo effect, the AC Josephson effect is strongly reduced. This result is compared to NRG theory and is attributed to a transition between the singlet ground state and the doublet excited state which is  enabled only when the junction is driven out-of-equilibrium by a voltage bias.
\end{abstract}

\maketitle

The AC Josephson effect \cite{Josephson1962} is the phenomenon by which a superconducting weak link with a bias voltage $V$ generates an oscillating current at frequency  $\nu_J=2eV/h$. It has been used to explore the Andreev Bound States (ABS) spectrum, which determines the supercurrent carried by the junction. For instance, its measurement points towards the topologically protected crossing of ABS in HgTe \cite{Wiedenmann2016,Deacon2017}, InAs nanowires \cite{Laroche2019} and Dirac semi-metals \cite{Li2018}. However, when probing the AC Josephson effect, due to the applied voltage bias and the resulting time evolution of the superconducting phase, the junction is driven out-of-equilibrium. This changes the occupation of the Andreev levels and the dynamics of the quasi-particles (QP) in the system \cite{Billangeon2007, Basset2019} and can lead to new physical effects not accessible at equilibrium. 

In the present article, we explore this out-of-equilibrium situation in a Josephson junction based on a carbon nanotube (CNT) quantum dot (QD) in the Kondo regime. In such a junction, without Kondo effect, electron-electron interaction results in Coulomb blockade, which gives rise to a doublet spin $1/2$ state if there is an odd number of electrons on the dot. In this doublet state, the Cooper pairs pass through the QD thanks to sequential cotunelling processes that involve a spin flip. This manifests as a reduced critical current and a $\pi$ shift of the current-phase relation \cite{Vandam2006,Cleuziou2006,Jorgensen2007,Francesci2010}, called $\pi$-junction. However, due to the coupling to the reservoirs and local interactions, Kondo correlations can develop. 
The Kondo effect is a many-body interaction between a localized impurity spin and free conduction electrons leading to the screening of the impurity spin \cite{Pustilnik2004,Goldhaber-Gordon1998,Cronenwett1998}. By enhancing cotunelling processes, a Kondo resonant state arises that is able to overcome the Coulomb blockade by opening a perfectly transmitted channel through the quantum dot when connected to normal reservoirs. For superconducting reservoirs, if the Kondo temperature  is larger than the superconducting gap ($k_BT_K>\Delta$), the Kondo effect and the superconductivity cooperate and enhance the DC Josephson effect by restoring the spin 0 singlet state (0-junction behavior).  

This singlet to doublet transition in QDs has attracted a large theoretical interest (see the recent review \cite{Meden2019}). Experimentally, it is now well established that it can be driven by the gate voltage \cite{Francesci2010,Maurand2012}, the magnetic field \cite{Garcia2020} and the superconducting phase \cite{Delagrange2015,Delagrange2016}. Here, we explore this transition by using the measurement of the Josephson emission as a probe of the state of the junction and find a range of parameters where this transition is forbidden at equilibrium  but  enabled  when the junction is driven out-of-equilibrium by a voltage bias.

\begin{figure}[tpb]
\begin{center}
   \includegraphics[width=8.5cm]{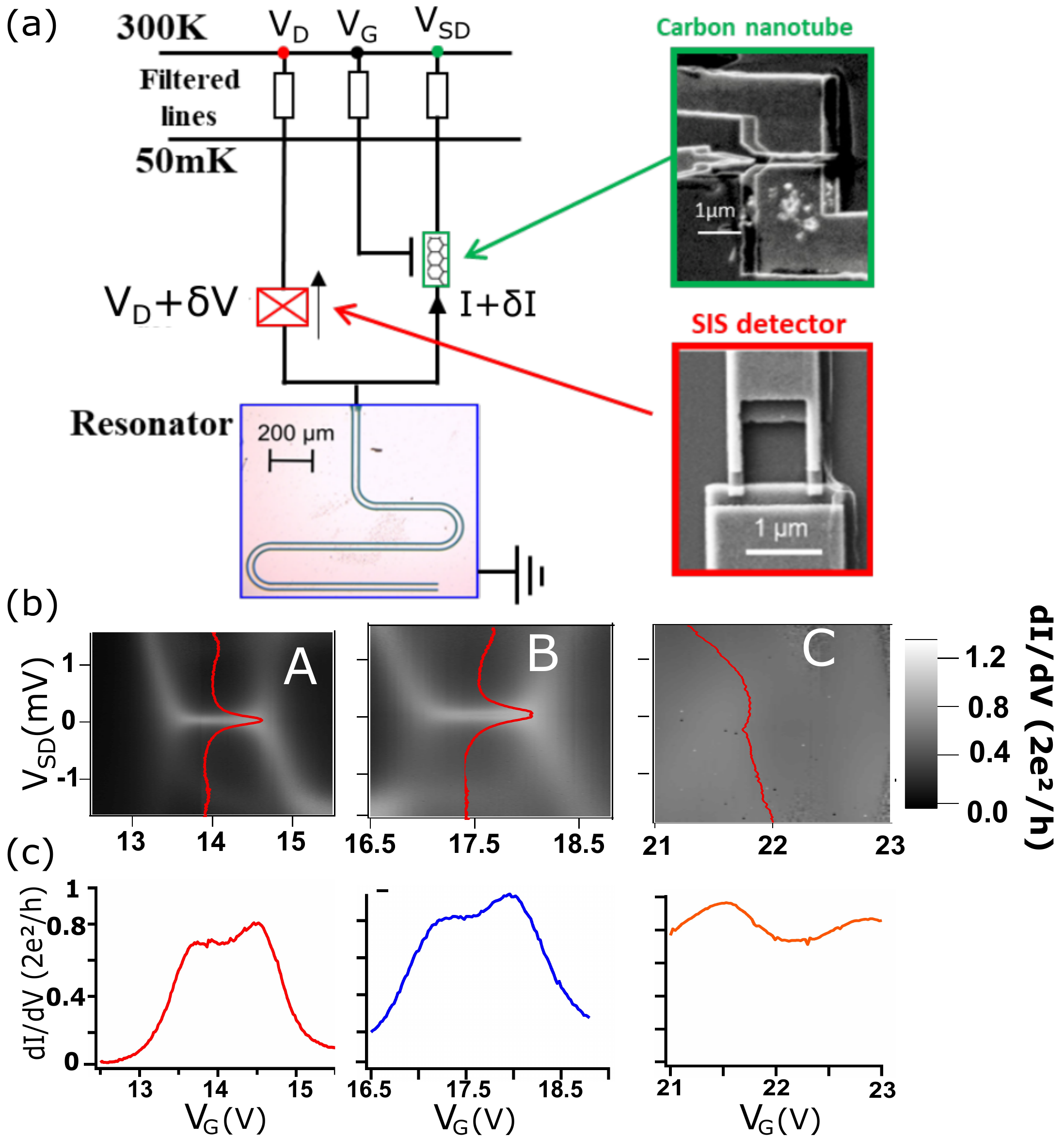}
    \end{center}
    \caption{Sample and normal state characterization. (a) The CNT Josephson junction is coupled to a SIS junction via a coplanar waveguide resonator. (b) Differential conductance $dI/dV_{SD}$ as a function of bias voltage $V_{SD}$ and gate voltage $V_G$ for Kondo ridge A and B and zone C. $dI/dV_{SD}$ curves taken at gate voltages 14 V, 17.5 V and 21.5 V are shown on top of the colorplot (red curves). (c) Conductance at $V_{SD}=0$ of Kondo ridges A and B, and for zone C.}
    \label{fig1}
\end{figure}

The Josephson emission of the CNT-based Josephson junction is probed using an on-chip detection scheme. In the gate regions that exhibit Kondo features in the normal state, the supercurrent is found to be enhanced by the Kondo effect whereas the Josephson emission is found to be strongly reduced such that the AC emission is not proportional to the DC supercurrent, as it could be naively expected from the Josephson relations. This striking result strongly suggests a dynamical change in the state of the junction, from singlet to doublet, induced by the phase evolution of the junction and the quasiparticle dynamics in the QD. 

The CNT Josephson junction is coupled to an on-chip detector, a superconductor-insulator-superconductor (SIS) junction \cite{Deblock2003,Basset2010,Delagrange2018}, via a coplanar waveguide resonator (Fig. \ref{fig1}a). The CNTs are grown by chemical vapor deposition on an oxidized undoped silicon substrate \cite{kasumov2007}. The contacts to the tube, the detector and the resonant circuit are made using electron beam lithography and metal deposition. The contacts are 400 nm apart and made of Pd (7nm)/ Al (100nm) ($\Delta=50 \mu eV$) or Pd(8nm)/Nb(11nm)/Al(50nm) ($\Delta=150 \mu eV$). Pd provides good contact to the CNTs, however reduces $\Delta$ compared to bare Al or Nb. Superconductivity in the Pd/Al contact is suppressed by a low magnetic field of 0.1T, without affecting the normal state of the CNT, allowing a good determination of the parameters of the dots. However for the Pd/Nb/Al contacts, the needed magnetic field is at least 1T, thus preventing a reliable extraction of all the dot parameters. The sample is cooled down in a dilution refrigerator of base temperature 50 mK and measured through low-pass filtered lines. The differential conductance is probed with a lock-in technique.

\begin{figure}[tpb]
     \begin{center}
     \includegraphics[width=8.5cm]{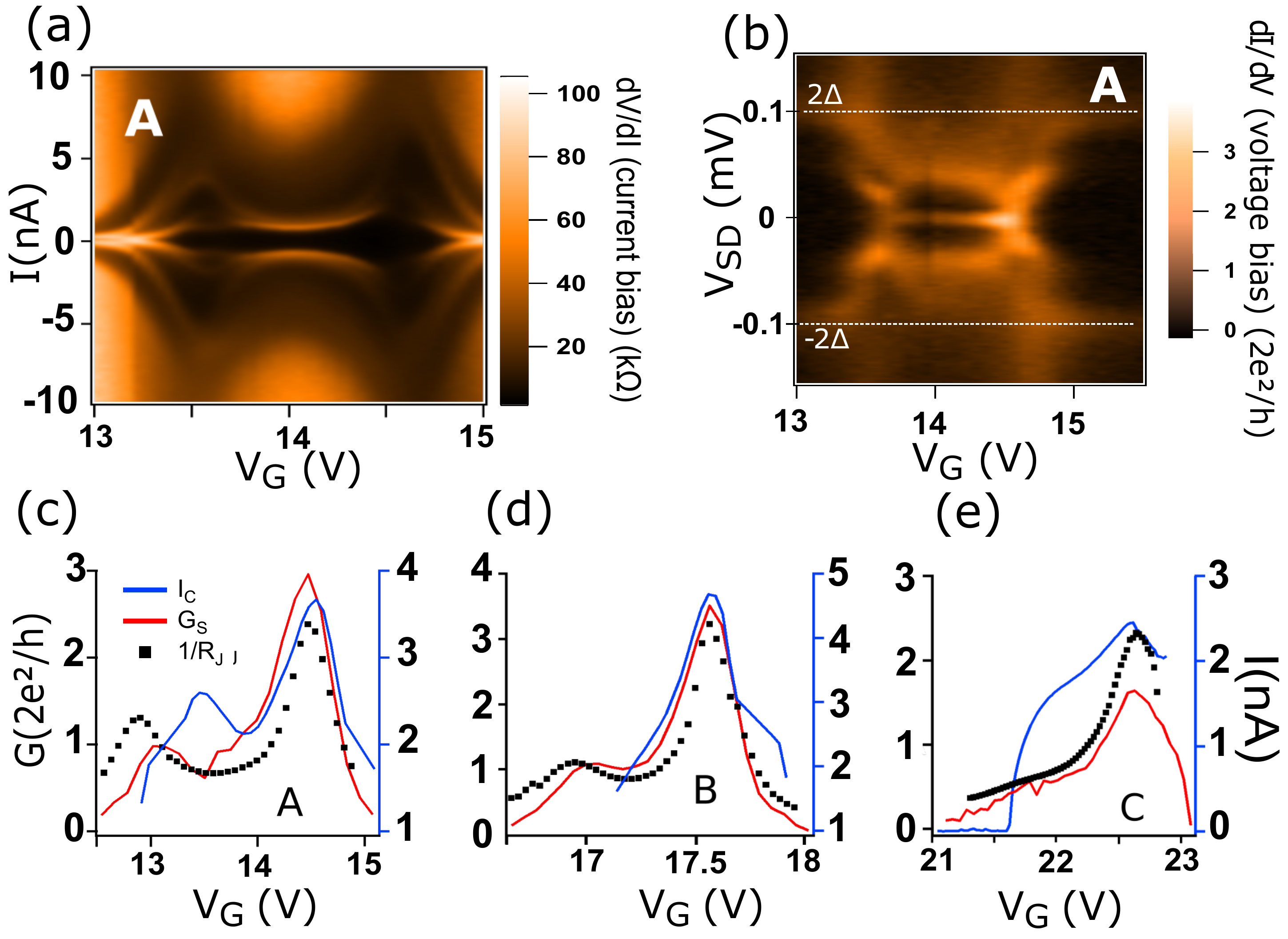}
     \end{center}
     \caption{DC Josephson effect. (a) Differential resistance $dV_{SD}/dI$ of the CNT Josephson junction as a function of the bias current $I$ and $V_G$ for Kondo ridge A. (b) Differential conductance $dI/dV_{SD}$ of the CNT Josephson junction as a function of $V_{SD}$ and $V_G$ for Kondo ridge A. (c)-(e) Gate dependence for region A, B and C of the critical current $I_C$ (blue curve) and the inverse of the resistance $R_J$ (square dots) extracted from the RCSJ model (see text and \cite{SM}). $G_S=dI/dV_{SD}$ at zero bias is plotted as the red line.}
     \label{fig2}
\end{figure} 

When biased by a voltage $V_{SD}$, the CNT emits photons at the Josephson frequency $\nu_J = 2eV_{SD}/h$. The SIS detector absorbs those photons, inducing a photo-assisted tunneling (PAT) current $I_{PAT}$. The resonant coupling circuit between detector and CNT transmits only at the resonance frequencies, designed to be $\nu_0=12.5$ GHz and odd harmonics. 
The $I_{PAT}$ current through the detector, biased at $V_D$ such that $2\Delta -h\nu < V_D < 2\Delta$, is proportional to the amplitude of the AC Josephson emission $I_C^{AC}$, following the relation \cite{Basset2010}:
\begin{equation}
I_{PAT}= \frac{1}{(2 V_{SD})^2} \frac{(I_C^{AC})^2}{4} |Z_t(2eV_{SD}/h)|^2 I_{qp}^0(V_D+2V_{SD})
\label{Ipat}
\end{equation}                                                      
with $I_{qp}^0(V_D)$ the $IV$ characteristic of the detector without irradiation and $Z_t(\nu)$ the impedance of the resonant circuit at frequency $\nu$. $I_{PAT}$ is obtained experimentally as the DC current in the SIS detector, and allows to extract $I_C^{AC}$ using equation (\ref{Ipat}) (see SM). 

\begin{figure}[tpb]
     \begin{center}
     \includegraphics[width=8.5cm]{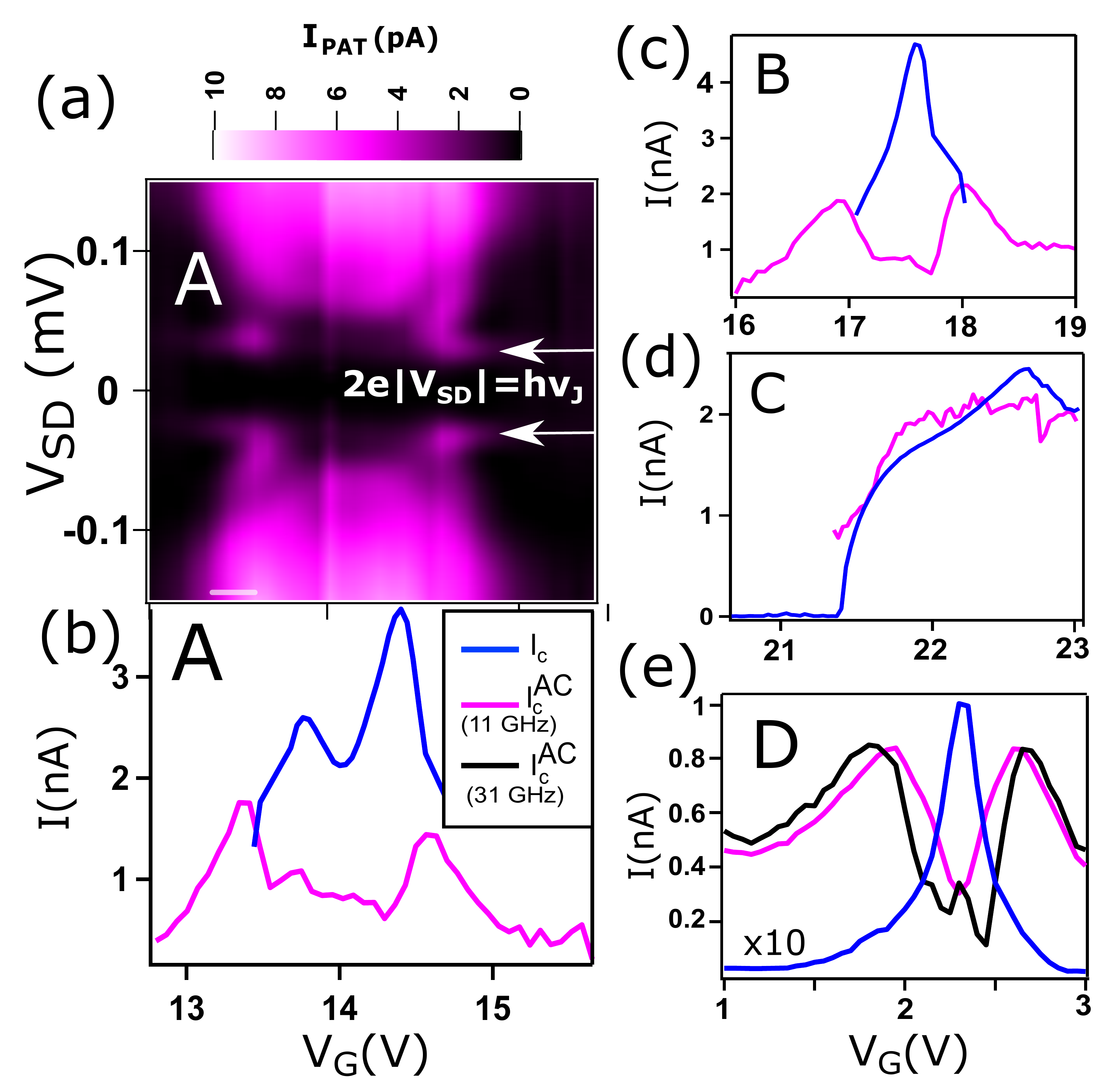}
     \end{center}
     \caption{Josephson emission. (a) PAT current in the detector as a function of $V_{SD}$ and $V_G$ for zone A. (b-d) Comparison of the critical current $I_C$ and the AC critical current $I_C^{AC}$ for zone A, B and C.  (e) Same comparison for the sample with Pd/Nb/Al contact and two resonator modes ($\nu_0=11$ GHz and $\nu_1=31$ GHz). The curve for the critical current has to be multiplied by 10. }
     \label{fig3}
\end{figure}

The CNT with Pd/Al contacts is characterized first in the normal state. The differential conductance $dI/dV_{SD}$ of the CNT is measured as a function of the bias voltage ($V_{SD}$) and the gate voltage ($V_G$) (see Fig \ref{fig1}b). The stability diagram of the QD exhibits Coulomb blockade diamonds with the four-fold degeneracy found for clean CNT QDs. In the diamonds with odd number of electrons, the Kondo effect manifests through a high conductance region at zero bias, the Kondo ridge. We focus on two Kondo ridges A and B, with filling N=1 and N=3, respectively. On figure \ref{fig1}b-c we also show another gate region, called hereafter region C, with a conductance close to the conductance quantum but without Kondo features. We consider as well a similar Kondo ridge in the sample with Pd/Nb/Al contacts, that we call region D \cite{SM}. 
The different parameters of the QD (see Table \ref{table1}), described with the single impurity Anderson model, are extracted for the Kondo ridges A and B. The charging energy $U$ is deduced from the size of the Coulomb diamond, the asymmetry $a=\Gamma_L/\Gamma_R$ of the contact from the value of the zero-bias conductance at the particle-hole symmetry point and the coupling to the reservoirs $\Gamma=\Gamma_L+\Gamma_R$ from the gate dependence of the Kondo temperature $T_K$ \cite{SM}. 
\begin{table}
	\centering
	\begin{tabular}{|c|c|c|c|c|c|}
		\hline
 		& $T_K(K)$ & $U$ (meV) & $\Gamma$ (meV) & a & $\Delta$ (meV) \\
 		\hline
 		Kondo A & 1.1 & 3.9 & 0.62 & 3.3 & 0.05 \\
 		Kondo B & 1.7 & 4 & 0.75 & 2.5 & 0.05 \\
 		Kondo D & $>\Delta $ & 2 &  &  & 0.15\\
 		\hline
	\end{tabular}
	\caption{Parameters of the CNT QD in the three Kondo regions considered in this letter. Because of the high critical field of the Pd/Nb/Al contacts, the Kondo region D could not be fully characterized.}
	\label{table1}
\end{table}

We now turn to the superconducting regime where a supercurrent can flow through the junction. The differential resistance $dV_{SD}/dI$ is measured as a function of the bias current $I$ and $V_G$ (Fig. \ref{fig2}a). We use the resistively and capacitively shunted junction (RCSJ) model \cite{novotny2005,Jorgensen2007,Eichler2009} to model the electromagnetic environment of the junction and account for a dissipative Josephson branch \cite{SM}. We extract from this model the value of the critical current $I_c$ and the resistance of the junction $R_J$ (Fig. \ref{fig2}b-d). The three regions A, B and C exhibit a gate modulated critical current and resistance $R_J$. $1/R_J$ behaves quite similarly to the measured conductance in the superconducting regime $G_S$. The fact that the critical current in the Kondo regions A and B remains relatively large is a good indicator that the QD stays in the singlet state, leading to a 0 junction behaviour. This is expected from the ratio $k_B T_K/\Delta$  and the asymmetry \cite{Kadlecova2017,Kadlecova2019}. The same qualitative behaviour is seen on sample with Pd/Nb/Al contacts, where the Josephson branch is less dissipative and allows direct extraction of $I_c$, without using the RCSJ model \cite{SM}. 

\begin{figure}[tpb]
     \begin{center}
     \includegraphics[width=8.5cm]{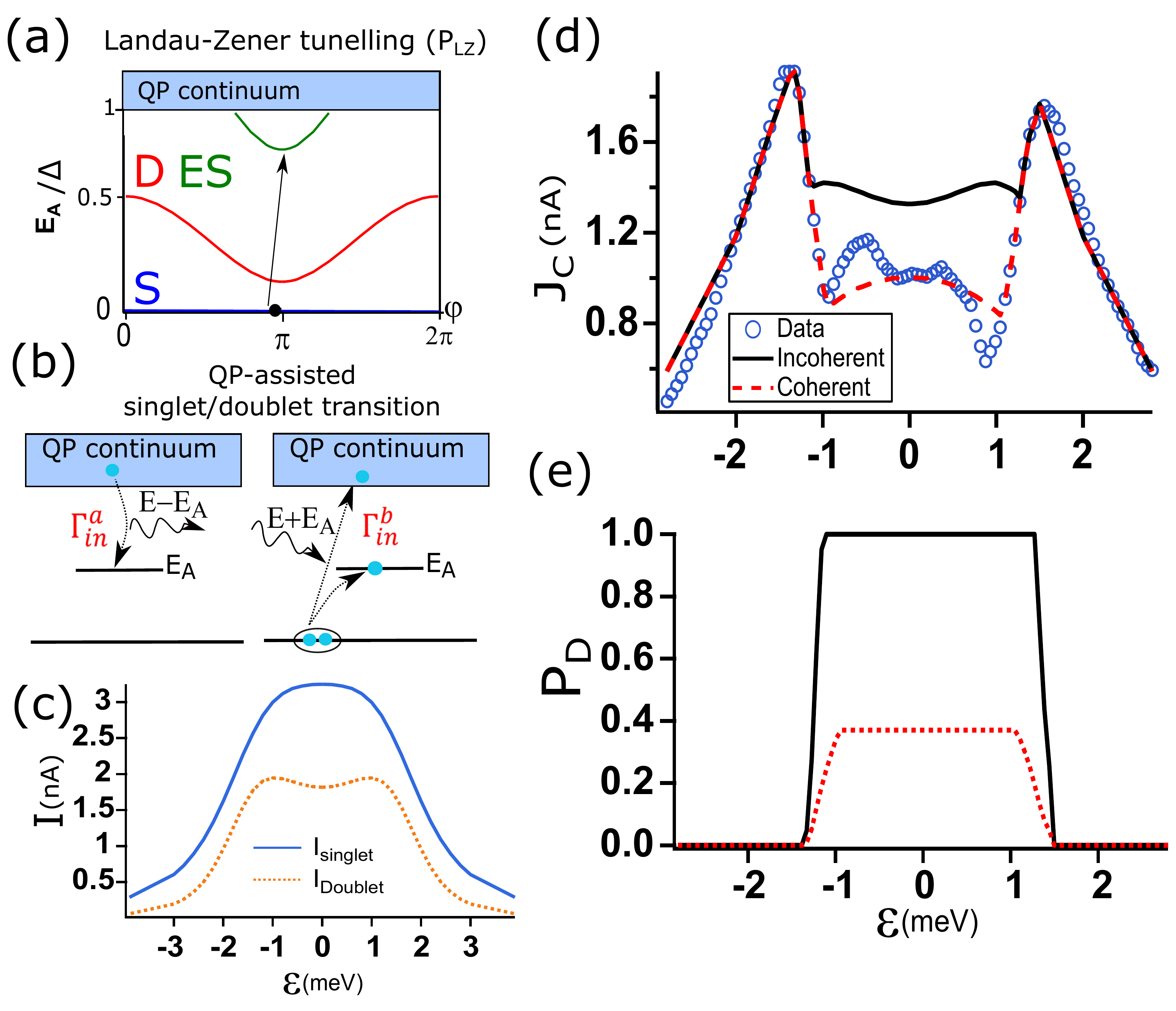}
     \end{center}
     \caption{(a) NRG many-body spectrum as a function of the superconducting phase $\varphi$ at the particle-hole symmetry point for the Kondo ridge A. The ground state is singlet (S), the first excited state a doublet (D, red curve) and the second excited state a singlet (ES, green curve). The arrow illustrates the Landau-Zener process, where the system tunnels from S to ES. (b) Illustration of two QP-assisted processes to excite the doublet state (D), associated with energy exchange with the environment. (c) Calculated amplitude of the first harmonics of the current phase relation for zone A in the singlet and doublet state. (d) Comparison between the data for Kondo ridge A (blue circles) and the calculated amplitude of the AC supercurrent $I_C^{AC}$ introducing a finite probability $P_D$ for the system to be in the doublet state using incoherent (black solid line) and coherent scenario (red dashed line) (see text) (e) $P_D$ in the incoherent (black solid line) and coherent regime (red dashed line).}
     \label{fig4}
\end{figure}

We have performed numerical renormalization group (NRG) calculations \cite{Zitko2009,Zitko2014} of the energy spectrum and supercurrent (Fig. \ref{fig4}a,c) using the parameters determined in the normal state (table \ref{table1}). They confirm that the ground state of the system for region A and B is always the singlet state. This leads to a supercurrent in the nanoampere range, consistent with the experiment, with the phase behaviour of a 0-junction.

To measure the Josephson emission, we voltage bias the detector such that $2\Delta-h\nu_{0}<|V_D|<2\Delta$ and measure simultaneously $dI/dV_{SD}$ (fig. \ref{fig2}b) and the PAT current (fig. \ref{fig3}a) as a function of $V_G$ and $V_{SD}$. On fig. \ref{fig2}b, the conductance of the sample shows the onset of QP tunneling at $V_{SD}=\pm 0.1$mV,  corresponding to $\Delta=50 \mu$eV. There is a strong increase of conductance at zero bias due to the supercurrent branch. Below the superconducting gap, finite conductance features are related to multiple Andreev reflection (MAR) processes. 

On fig. \ref{fig3}a, the PAT current reveals that the emission of the CNT junction has two contributions. One is the AC Josephson effect of the junction, at the Josephson frequency $\nu_J=2eV_{SD}/h$. The second contribution is broadband and associated to MAR processes and QP tunneling. In the PAT response, we do not detect any signature of harmonics in the AC Josephson effect (expected at voltage $V_{SD}=h\nu_J/2 n e$ for the $n^{th}$ harmonics). Consequently we separate the two processes by attributing the peak at $\nu_J$ to the AC Josephson effect and the remaining baseline to the broadband contribution (see Fig. 4 of \cite{SM}). On figure \ref{fig3}b-e the amplitude of the dynamical critical current $I_C^{AC}$ is plotted, extracted using formula (\ref{Ipat}) from the peak at $V_{SD}=h\nu/(2e)$ in the PAT current. 
In the reference region C, where there is neither Coulomb blockade nor Kondo effect, this dynamical critical current follows the DC critical current $I_C$. By contrast close to the center of the Kondo regions A, B and D, there is a strong reduction of $I_C^{AC}$, in a region where the critical current $I_C$ is enhanced due to the Kondo correlations (Fig. \ref{fig3}b-e). In region D, this effect could be observed as well at the 31 GHz mode of the resonator, and is even enhanced compared to the fundamental frequency (Fig. \ref{fig3}e). This collapse of the AC Josephson emission, enhanced as the Josephson frequency increases and specific to the Kondo regions, is the central result of this work.

We now turn to possible explanations for the reduction of the AC Josephson effect in the Kondo regime. One may think about the decoherence of the Kondo effect due to voltage induced spin relaxation \cite{muller2013,basset2012,kaminski1999,kaminski2000,paaske2004}, or dynamical effects similar to the high-frequency cutoff for the emission of a quantum dot in the normal state \cite{Delagrange2018}. However, because $e V_{SD}/k_B T_K <1$, $h\nu_J /k_B T_K <1$ and the asymmetry of the coupling to the contacts, these effects shall remain small.  

Another possible process is Landau-Zener (LZ) tunneling, that induces a transition to excited levels with
a probability which increases when the phase velocity (and thus the Josephson frequency) is high. This is what happens for a quantum channel junction with high transparency (\cite{Averin1995,SM}) and involves transitions between
singlet states, due to parity constrains \cite{Beenaker1991}. However, for the quantum dot, the energy of the excited singlet state is rather close to $\Delta$ (see fig. \ref{fig4}a). The probability $P_{LZ}$ of this transition, while not completely negligible ($P_{LZ} = 0.43$), depends little on $V_G$ \cite{SM}. This would not account for the suppression of the AC Josephson emission observed on fig \ref{fig3}.

The collapse of the Josephson emission is more likely related to transitions from the ground state singlet to the doublet state. This is made possible by the tunneling of QP in the QD thanks to energy exchange with the environment, also called QP poisoning (see fig. \ref{fig4}b). We first consider the situation where no bias voltage is applied, and find that the probability $P_D$ for the QD to be in the doublet state is small in a DC current configuration with a current bias varying in the kHz range (\cite{Olivares2014,SM}). This explains why the measurement of the DC critical current, which happens at the maximum of the supercurrent around $\varphi=\pi/2$, is consistent with a 0-junction and a singlet ground state. 

The situation can be quite different when one measures the AC emission. The theory for QD in this regime is not available, except for particular limits \cite{Hiltscher2012,Dellanna2008,Lamic2020}, and we give the following qualitative arguments. Due to the applied bias needed to have Josephson emission, the QP injection rate $\Gamma_{in}$ increases significantly compared to equilibrium. Moreover, close to the particle-hole symmetry point, the doublet state is detached from the continuum with a relatively large gap $\Delta_{cont}$. This keeps the escape rate of QPs relatively low, while the injection rate is increased, such that the probability for the QD to be in the doublet state is expected to be higher in a voltage biased situation. This leads to a decrease of $I_{AC}^C$ since the critical current of the doublet state is lower than the one of the singlet ground state. Despite a higher gap value, the samples with Pd/Nb/Al contacts exhibit the same phenomenon (Fig. \ref{fig3}e). This can be related to the existence of a soft gap for these samples, inducing a small but finite QP density at energy below the gap. 

Moving away from the electron-hole symmetry point, $\Delta_{cont}$ is reduced significantly (fig. 11 of \cite{SM}), which increases the probability for a QP present on the dot to escape thanks to Demkov Osherov tunneling processes between the doublet state and the continuum (\cite{Demkov1968, Houzet2013,Badiane2013,SM}. Concurrently the minimum value of energy of the doublet state, at $\varphi=\pi$, increases, which reduces the rate of QP injection into the QD. These two effects thus restore a high probability for the QD to be in the singlet ground state and increases its effective supercurrent. That is what is measured in the experiment. 

From the amplitude of the AC emission, it is possible to extract $P_D$ assuming that the dynamical Josephson current is given by $I_{AC}^C = P_D J_D + (1-P_D) J_S$, where $J_S$ is the amplitude of the critical current in the singlet state and $J_D$ the one in the doublet state (Fig. \ref{fig4}c). In an incoherent scenario, where the QP dynamics is not correlated with the phase evolution of the junction, only the absolute value of the amplitude of the singlet and doublet supercurrent is taken into account. With a probability one to be in the doublet state close to the particle-hole symmetry point, one can qualitatively reproduce the reduction of the supercurrent (Fig. \ref{fig4}d and e). Oppositely, in a coherent scenario, where the QP dynamics is correlated with the phase evolution of the junction, the sign of the supercurrent (positive for the singlet and negative for the doublet) has to be considered. This leads to a quantitative agreement with the data (Fig. \ref{fig4}d and e), with a finite probability to be in the doublet state, but puts a strong constrain on the model used to describe the dynamics of the junction. 

To conclude we have explored the singlet to doublet transition of an out-of-equilibrium CNT Josephson junction  by probing its Josephson emission. It is strikingly reduced in the gate regions where the critical current is enhanced due to the interplay of the Kondo effect and the superconducting proximity effect. We interpret this result as a transition between a singlet ground state and a doublet excited state induced by the dynamics of quasiparticles in the QD. This thus demonstrates the importance of taking into account electron-electron interactions and non-equilibrium processes to understand the dynamics of QD Josephson junctions. 

\textit{Acknowledgments}: The authors acknowledge T.~Novotn\'y, M. Houzet, J. Meyer, S. Gu\'eron, J. Basset, M. Aprili and P. Simon for fruitful discussions and technical help from S. Autier-Laurent and R. Weil. This work was supported by Grant No. 19-13525S of the Czech Science Foundation (AK), the French program ANR JETS (ANR-16-CE30-0029-01) and the Region Ile-de-France in the framework of DIM SIRTEQ . Computational resources were supplied by the project "e-Infrastruktura CZ" (e-INFRA LM2018140) provided within the program Projects of Large Research, Development and Innovations Infrastructures.


\begin{thebibliography}{10}

\bibitem{Josephson1962}
B. D. Josephson, Phys. Lett. \textbf{1} (7), 251–253 (1962). 

\bibitem{Wiedenmann2016}
J. Wiedenmann et al., Nature Communications \textbf{7}, 10303 (2016).

\bibitem{Deacon2017} 
R. S. Deacon et al., Phys. Rev. X \textbf{7}, 021011 (2017).

\bibitem{Laroche2019}
D. Laroche et al., Nature Communications \textbf{10}, 245 (2019).

\bibitem{Li2018}
C. Li, J. C. de Boer, B. de Ronde, S. V. Ramankutty, E. van Heumen, Y. Huang, A. de Visser, A. A. Golubov, M. S. Golden and A. Brinkman, {\em Nature Materials} \textbf{17}, 875-880 (2018).

\bibitem{Basset2019} 
J. Basset, M. Kuzmanovic, P. Virtanen, T. T. Heikkilä, J. Estève, J. Gabelli, C. Strunk, and M. Aprili, {\em Phys. Rev. Research} \textbf{1}, 032009 (2019).

\bibitem{Billangeon2007}
P.-M. Billangeon, F. Pierre, H. Bouchiat, R. Deblock, {\em Phys. Rev. Lett.} \textbf{98}, 216802 (2007).

\bibitem{Vandam2006}
J.~A. van Dam, Y.~V. Nazarov, Erik P. A.~M. Bakkers, S.
De~Franceschi, and L.~P. Kouwenhoven, 
\newblock {\em Nature} \textbf{442,} 667 (2006).

\bibitem{Cleuziou2006}
J.-P. Cleuziou, W.~Wernsdorfer, V.~Bouchiat, T.~Ondar{\c c}uhu, and
M.~Monthioux.
\newblock {\em Nature Nanotechnology} \textbf{1,} 53 (2006).

\bibitem{Jorgensen2007}
H. I. J$\phi$rgensen, T. Novotn\'y, K. Grove-Rasmussen, K. Flensberg, and P. E. Lindelof, {\em Nano Lett.} \textbf{7}, 2441 (2007).

\bibitem{Francesci2010}
S. De Franceschi, L. Kouwenhoven, C. Schönenberger, and W. Wernsdorfer, Nature Nanotechnology \textbf{5}, 703 (2010).

\bibitem{Pustilnik2004}
M. Pustilnik and L. Glazman,
\newblock {\em Journal of Physics: Condensed Matter} \textbf{16,} R513 (2004).

\bibitem{Goldhaber-Gordon1998}
D. Goldhaber-Gordon, J. G\"ores, M.~A. Kastner, H. Shtrikman, D. Mahalu and U.
Meirav, {\em Phys. Rev. Lett.} \textbf{81,} 5225-5228 (1998).

\bibitem{Cronenwett1998}
S. M. Cronenwett, T. H. Oosterkamp, and L. P. Kouwenhoven,
\newblock {\em Science} \textbf{281,} 540--544 (1998).

\bibitem{Meden2019}
V. Meden, J. Phys.: Condens. Matter \textbf{31}, 163001 (2019).	
	
\bibitem{Maurand2012}
R. Maurand, T. Meng, E. Bonet, S. Florens, L. Marty, and W. Wernsdorfer.
\newblock {\em Phys. Rev. X}, \textbf{2,} 019901 (2012).

\bibitem{Garcia2020}
A. Garc\'ia Corral, D. M. T. van Zanten, K. J. Franke, H. Courtois, S. Florens, and C. B. Winkelmann, {\em Phys. Rev. Research} \textbf{2}, 012065 (2020).

\bibitem{Delagrange2015}
R. Delagrange, D. J. Luitz, R. Weil, A. Kasumov, V. Meden, H. Bouchiat, and R. Deblock, {\em Phys. Rev. B} \textbf{91}, 241401 (2015).

\bibitem{Delagrange2016}
R. Delagrange, R. Weil, A. Kasumov, M. Ferrier, H. Bouchiat, and R. Deblock, {\em Phys. Rev. B} \textbf{93}, 195437 (2016) .

\bibitem{Deblock2003}
R. Deblock, E. Onac, L. Gurevich, and L.P. Kouwenhoven, {\em Science} \textbf{301}, 203 (2003).

\bibitem{Delagrange2018}
R. Delagrange, J. Basset, H. Bouchiat, and R. Deblock, {\em Phys. Rev. B} \textbf{97}, 041412(R) (2018).

\bibitem{Basset2010} 
J. Basset, H. Bouchiat, and R. Deblock, {\em Phys. Rev. Lett.} \textbf{105}, 166801 (2010).

\bibitem{kasumov2007}
Y. Kasumov et al., {\em Appl. Phys. A} \textbf{88}, 4 (2007), pp. 687-691.

\bibitem{novotny2005}
T. Novotn\'y, A. Rossini, and K. Flensberg, {\em Phys. Rev. B} \textbf{72}, 224502 (2005).

\bibitem{Eichler2009}
A.~Eichler, R.~Deblock, M.~Weiss, C.~Karrasch, V.~Meden, C.~Sch\"onenberger, and H.~Bouchiat.
{\em Phys. Rev. B}, \textbf{79,} 161407 (2009).

\bibitem{Kadlecova2017}
A. Kadlecov\'a, M. \v{Z}onda, and T.~Novotn\'y, {\em Phys. Rev. B} \textbf{95}, 195114 (2017).

\bibitem{Kadlecova2019}
A. Kadlecov\'a, M. \v{Z}onda, V. Pokorn\'y, T.~Novotn\'y, {\em Phys. Rev. Applied} \textbf{11}, 044094 (2019).

\bibitem{Zitko2014}
R. \v{Z}itko, NRG Ljubljana - open source numerical renormalization group code (2014), http://nrgljubljana.ijs.si

\bibitem{Zitko2009}
R. \v{Z}itko and T. Pruschke, {\em Phys. Rev. B} \textbf{79}, 085106 (2009).

\bibitem{muller2013}
S. Y. M\"uller, M. Pletyukhov, D. Schuricht, and S. Andergassen, {\em Phys. Rev. B} \textbf{87}, 245115 (2013).

\bibitem{basset2012}
J. Basset, A. Y. Kasumov, C. P. Moca, G. Zar\`and, P. Simon, H. Bouchiat, and R. Deblock, {\em Phys. Rev. Lett.} \textbf{108}, 046802 (2012).

\bibitem{kaminski1999}
A. Kaminski, Y. V. Nazarov, and L. I. Glazman, {\em Phys. Rev. Lett.} \textbf{83}, 384 (1999).

\bibitem{kaminski2000}
A. Kaminski, Y. V. Nazarov, and L. I. Glazman, {\em Phys. Rev. B} \textbf{62}, 8154 (2000).

\bibitem{paaske2004}
J. Paaske, A. Rosch, J. Kroha, and P. W\"olfle, {\em Phys. Rev. B} \textbf{70}, 155301 (2004).

\bibitem{Averin1995}
D. Averin and A. Bardas, {\em Phys. Rev. Lett.} \textbf{75}, 1831 (1995).

\bibitem{Beenaker1991}
C. W. J. Beenakker and H. van Houten, {\em Phys. Rev. Lett.} \textbf{66}, 3056 (1991).

\bibitem{Olivares2014}
D. G. Olivares, A. L. Yeyati, L. Bretheau, C.O. Girit, H. Pothier, and C. Urbina, {\em Phys. Rev. B} \textbf{89}, 104504 (2014).

\bibitem{Hiltscher2012}
B. Hiltscher, M. Governale, and J. K\"onig, {\em Phys. Rev. B} \textbf{86}, 235427 (2012).

\bibitem{Dellanna2008}
L. Dell’Anna, A. Zazunov, and R. Egger, {\em Phys. Rev. B} \textbf{77}, 104525 (2008).

\bibitem{Lamic2020} B. Lamic, J. S. Meyer, and M. Houzet, {\em Phys. Rev. Research} \textbf{2}, 033158 (2020).

\bibitem{Demkov1968}
Y.N. Demkov, V.I. Osherov, {\em Sov. Phys. JETP} \textbf{26}, 916 (1968).

\bibitem{Houzet2013}
M. Houzet, J. S. Meyer, D. M. Badiane, and L. I. Glazman, {\em Phys. Rev. Lett.} \textbf{111}, 046401 (2013).

% REFERENCES OF THE SM
\bibitem{SM}
See Supplemental Material for a discussion of the extraction of Kondo temperature and parameters of the quantum dot, the critical current, the measurement of the PAT current and AC emission, a presentation of data for other samples, details on NRG calculation of the Andreev spectrum and the supercurrent, a comparison of quantum dot and quantum channel Josephson junctions and finally the evaluation of the quasiparticle dynamics in the QD junction. It includes Refs. \cite{Tsvelick1983,Bickers1987,Babic2004,Jespersen2006,Kretinin2011,Costi1994,Sasaki2004,Delattre2009,Nygard2000,Ferrier2016,Laird2015,Hata2018,Saldana2019,Ambegaokar1963,Yeyati-03,Vecino-04,Mullen1988,Badiane2013,Cuevas-96}.

\bibitem{Tsvelick1983} A.M. Tsvelick and P.B. Wiegmann, {\em Adv. Phys.}
\textbf{32}, 453 (1983).

\bibitem{Bickers1987} N. E. Bickers, {\em Rev. Mod. Phys.} \textbf{59}, 845 (1987).

\bibitem{Babic2004} B. Babi\'c, T. Kontos, and C. Sch\"{o}nenberger,
{\em Phys. Rev. B} \textbf{70}, 235419 (2004).

\bibitem{Jespersen2006} T. S. Jespersen, M. Aagesen, C. S{\o}rensen,
P. E. Lindelof, and J. Nyg{\aa}rd, {\em Phys. Rev. B} \textbf{74}, 233304 (2006).

\bibitem{Kretinin2011} A. V. Kretinin, H. Shtrikman, D. Goldhaber-Gordon,
M. Hanl, A. Weichselbaum, J. von Delft, T. Costi, and D. Mahalu, {\em Phys.
Rev. B} \textbf{84}, 245316 (2011).

\bibitem{Costi1994} T. A. Costi, A. C. Hewson, and V. Zlatic, {\em J.
Phys.: Condens. Matter} \textbf{6}, 2519 (1994).

\bibitem{Sasaki2004}
S. Sasaki, S. Amaha, N. Asakawa, M. Eto, and S. Tarucha, Phys. Rev. Lett. \textbf{93}, 017205 (2004).

\bibitem{Delattre2009}
T. Delattre, C. Feuillet-Palma, L. G. Herrmann, P. Morfin, J.-M. Berroir, G. Fève, B. Plaçais, D. C. Glattli, M.-S. Choi, C. Mora, and T. Kontos, Nat. Phys. \textbf{5}, 208 (2009).

\bibitem{Nygard2000}
J. Nyg{\aa}rd, D. H. Cobden, and P. E. Lindelof, Nature \textbf{408}, 342 (2000).

\bibitem{Ferrier2016}
M. Ferrier, T. Arakawa, T. Hata, R. Fujiwara, R. Delagrange, R. Weil, R. Deblock, R. Sakano, A. Oguri, and K. Kobayashi, Nat. Phys. \textbf{12}, 230 (2016).

\bibitem{Laird2015} E. A. Laird, F. Kuemmeth, G. A. Steele, K. Grove-Rasmussen, J. Nyg{\aa}rd, K. Flensberg, and L. P. Kouwenhoven, Rev. Mod. Phys. \textbf{87}, 703 (2015).

\bibitem{Hata2018}
T. Hata, R. Delagrange, T. Arakawa, S. Lee, R. Deblock, H. Bouchiat, K. Kobayashi, and M. Ferrier, Phys. Rev. Lett. 121, 247703 (2018).

\bibitem{Saldana2019}
J. C. E. Saldaña, R. Žitko, J. P. Cleuziou, E. J. H. Lee, V. Zannier, D. Ercolani, L. Sorba, R. Aguado, and S. D. Franceschi, Science Advances 5, eaav1235 (2019).

\bibitem{Ambegaokar1963} V. Ambegaokar and A. Baratoff, {\em Phys. Rev.
Lett.} \textbf{10}, 486 (1963).

\bibitem{Yeyati-03} A. Levy-Yeyati, A. Mart\'in-Rodero, and E. Vecino, {\em Phys. Rev. Lett.} \textbf{91}, 266802 (2003).

\bibitem{Vecino-04} E. Vecino, M. R. Buitelaar, A. Mart\'in-Rodero, C. Schönenberger, and A. Levy-Yeyati, Solid State Communications \textbf{131}, 625 (2004).

\bibitem{Mullen1988} K. Mullen, Y. Gefen, and E. Ben-Jacob, {\em Physica
B: Condensed Matter} \textbf{152}, 172 (1988).

\bibitem{Badiane2013} D. M. Badiane, L. I. Glazman, M. Houzet, and
J. S. Meyer, {\em Comptes Rendus Physique} \textbf{14}, 840 (2013).

\bibitem{Cuevas-96} J. C. Cuevas, A. Mart\'in-Rodero, and A. Levy
Yeyati, {\em Phys. Rev. B} \textbf{54}, 7366 (1996).

\end{thebibliography}
\end{document}

\title{SUPPLEMENTAL MATERIAL \\ Collapse of the Josephson emission in a carbon nanotube junction in the Kondo regime}
\author{D. Watfa}
\affiliation{Universit\'e Paris-Saclay, CNRS, Laboratoire de Physique des Solides, 91405, Orsay, France.}
\author{R. Delagrange}
\affiliation{Universit\'e Paris-Saclay, CNRS, Laboratoire de Physique des Solides, 91405, Orsay, France.}
\author{A. Kadlecov\'a}
\affiliation{Department of Condensed Matter Physics, Faculty of Mathematics and Physics, Charles University, Ke Karlovu 5,
CZ-121 16 Praha 2, Czech Republic}
\author{M. Ferrier}
\affiliation{Universit\'e Paris-Saclay, CNRS, Laboratoire de Physique des Solides, 91405, Orsay, France.}
\author{A. Kasumov}
\affiliation{Universit\'e Paris-Saclay, CNRS, Laboratoire de Physique des Solides, 91405, Orsay, France.}
\author{H. Bouchiat}
\affiliation{Universit\'e Paris-Saclay, CNRS, Laboratoire de Physique des Solides, 91405, Orsay, France.}
\author{R. Deblock}
\affiliation{Universit\'e Paris-Saclay, CNRS, Laboratoire de Physique des Solides, 91405, Orsay, France.}

\maketitle

\section{Kondo temperature and parameters of the quantum dot}

The Kondo effect is possible only if the temperature is smaller than the Kondo temperature $T_K$. $T_K$ can be well approximated by the expression predicted by the Bethe Ansatz \cite{Tsvelick1983,Bickers1987}:
\begin{equation}
T_K=\sqrt{U\Gamma/2}\exp\left[- \frac{\pi}{8U
\Gamma}|4\epsilon^2-U ^2|\right] 
\label{formula_Tk}
\end{equation}
where $\epsilon$ is the energy shift measured from the center of the
Kondo ridge. Typically, in carbon nanotube quantum dots, the Kondo temperature reaches 1.5-2K \cite{Babic2004,Basset2012,Maurand2012}, values similar to the one obtained in InAs nanowires \cite{Jespersen2006,Kretinin2011}.

\begin{figure}[tb]
\begin{center}
   \includegraphics[width=8.6cm]{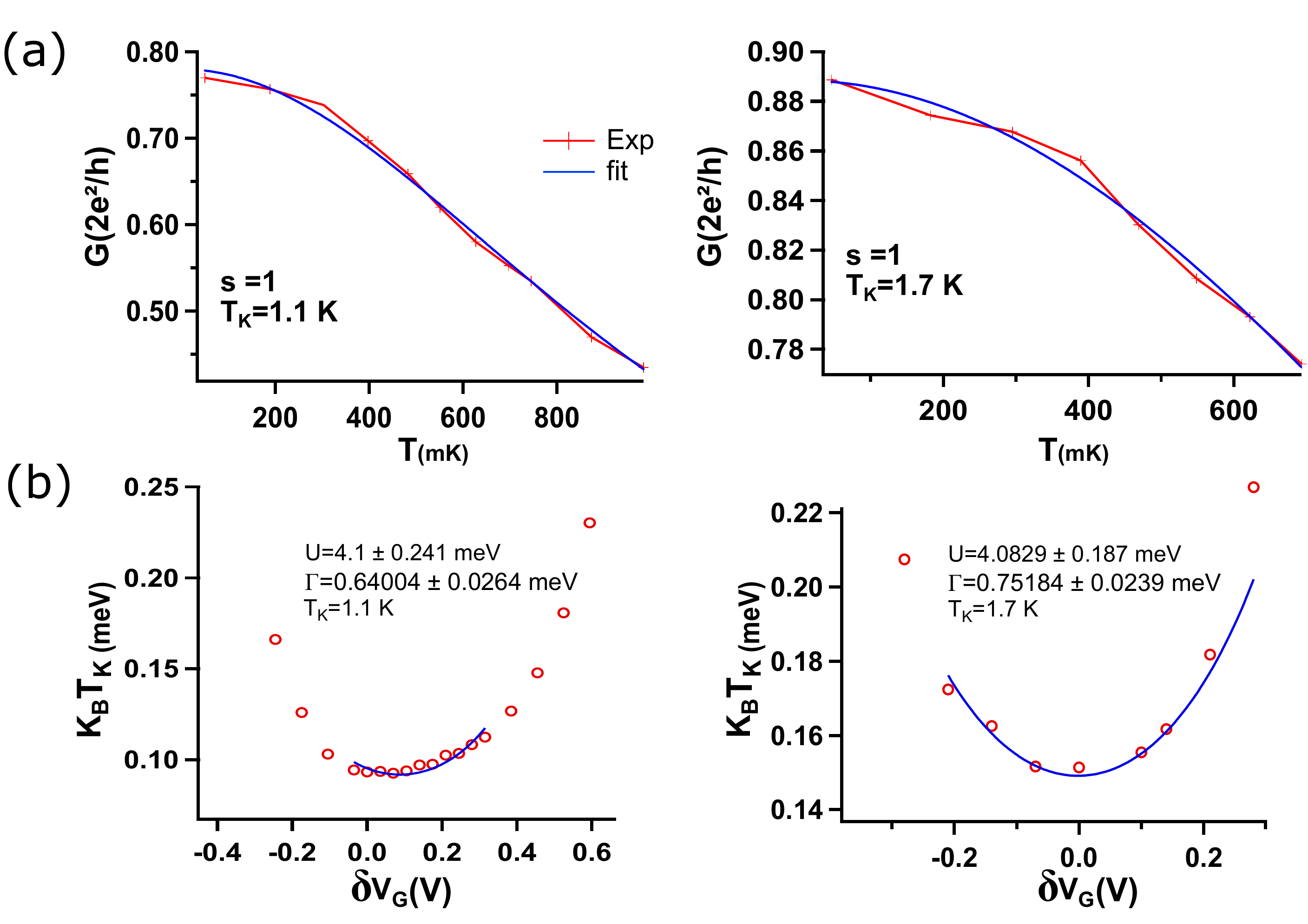}
    \end{center}
    \caption{(a) Temperature dependence of the zero-bias conductance of the CNT quantum dot for zone A and B. The agreement with the NRG calculation is not extremely good. As a consequence we took for the Kondo temperature the value of temperature where $G(T=T_K)=G_0/2$, with $G_0$ the value of the conductance at low temperature. (b) Gate dependence of the Kondo temperature extracted from the evolution of the width of the conductance peak as a function of bias voltage. The parameters extracted from the fit are shown in the legend of the figure.}
    \label{SM_fig1}
\end{figure}

A finite temperature of order $T_K$ results in a reduction of the conductance. The temperature dependence of the conductance can be described by the phenomenological expression :
\begin{equation}
\frac{dI}{dV}(T)=\frac{G_0}{(1+(2^{1/s}-1)(\frac{T}{T_K})^2)^s}
\end{equation}
with $G_0$ the conductance at very low temperature. NRG calculations have shown that spin-1/2 Kondo effect was best represented by $s=0.22$ \cite{Costi1994,Kretinin2011}. Fitting the conductance at zero-bias as a function of temperature allows one to extract the Kondo temperature (fig. \ref{SM_fig1}a). The agreement of this formula with our data is not completely satisfactory with $s=0.22$ and a better agreement is found taking $s=1$. Note that values of $s$ significantly higher than 0.22 have also been found in other experiments \cite{Sasaki2004,Delattre2009}. We define $T_K$ as the value of T where the conductance is divided by a factor 2. Note that this definition is independent of the parameter $s$. The value extracted this way is consistent with typical value found in carbon nanotube quantum dots in the SU2 regime \cite{Nygard2000,Eichler2009,Laird2015,Ferrier2016} and with the width (half width at half maximum) of the zero-bias conductance peak as a function of bias voltage $V_{SD}$.

\begin{figure}[tb]
\begin{center}
   \includegraphics[width=8.6cm]{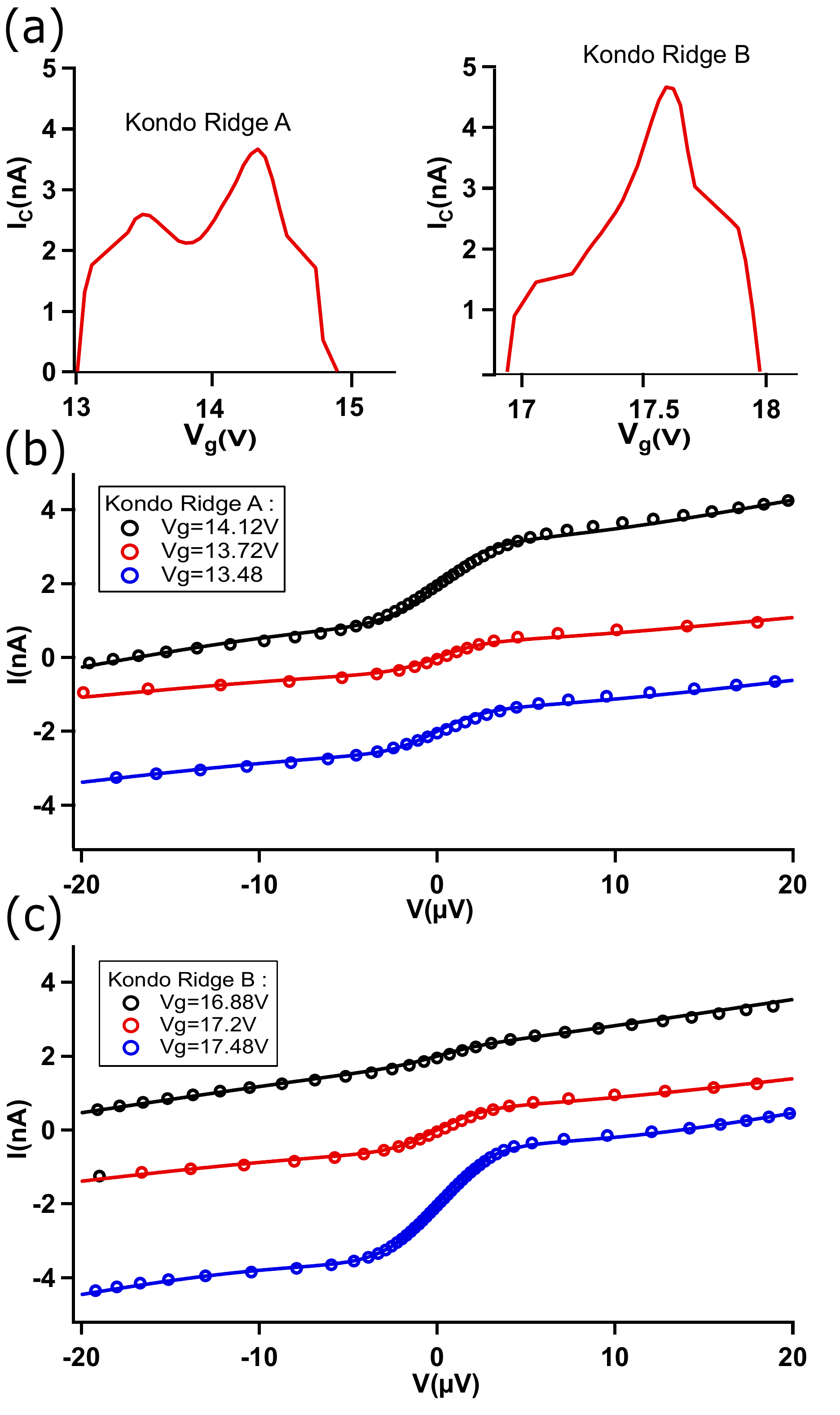}
    \end{center}
    \caption{(a) Gate dependence of the extracted critical current for Kondo ridge A and B. The parameters of the fit are $R=0.9$k$\Omega$ and $T=100$mK. (b) Data and theoretical curves for three gate voltages of the Kondo ridge A. (c) Data and theoretical curves for three gate voltages of the Kondo ridge B. }
    \label{SM_fig2}
\end{figure}

The gate dependence of the Kondo temperature (fig. \ref{SM_fig1}b) is then extracted from the width of the zero bias peak. By fitting these data by eq. \ref{formula_Tk} (a procedure which has been also used in Refs. \cite{Eichler2009,Kretinin2011,Ferrier2016,Garcia2020}), we can extract the parameters of the quantum dot. The charging energy is in agreement with the one deduced from the stability diagram. 

\section{Extraction of the critical current}

For the supercurrent measurement, the device is current
biased. We simultaneously use AC and DC bias while measuring the
resulting voltage drop across it. From the AC part, we obtain data
on the differential resistance. By numerical
integration we get I-V curves that show a supercurrent branch and a
smooth transition to a resistive branch with higher resistance. The transition between the two regimes is not hysteretic, and the supercurrent part exhibits a non zero resistance $R_S$ even at low bias. This behaviour is common in mesoscopic
Josephson junctions that have a high normal state resistance of the
order of the resistance quantum $h/e^2$. To extract the supercurrent, we
use a theory that explicitly includes the effect of the
dissipative electromagnetic environment in the framework of the extended RCSJ-model \cite{Jorgensen2007,Eichler2009,Hata2018,Saldana2019}.
The input parameters are the value of the external resistor $R$ 
and temperature $T$. The critical current $I_{c}$ and the junction resistance $R_{J}$ can then be extracted for every measured gate voltage (Fig. \ref{SM_fig2}(b) and (c)), from a fit to~:
\begin{equation}
I(V_{bias}) = \left\{ I_c  Im \left[
\frac{I_{1-i\eta}(I_c \hbar/2 e k_B
T)}{I_{-i\eta}(I_c \hbar/2 e k_B T)} \right]
 + \frac{V_{bias}}{R_j} \right\} \frac{R_j}{R_j+R} \label{IV}
\end{equation}
where $\eta=\hbar V_{bias}/2 e R k_B T$ and $I_{\alpha}(x)$ is the
modified Bessel function of complex order $\alpha$
\cite{Jorgensen2007}. The value of the critical current and the junction conductance $1/R_J$ are plotted on fig. \ref{SM_fig2}c-e of the main article. The parameters used for the fit are $R=0.9$k$\Omega$ and $T=100$mK.

\section{PAT current and AC emission}

\subsection{Calibration of the transimpedance}

\begin{figure}[tb]
    \begin{center}
    \includegraphics[width=8.6cm]{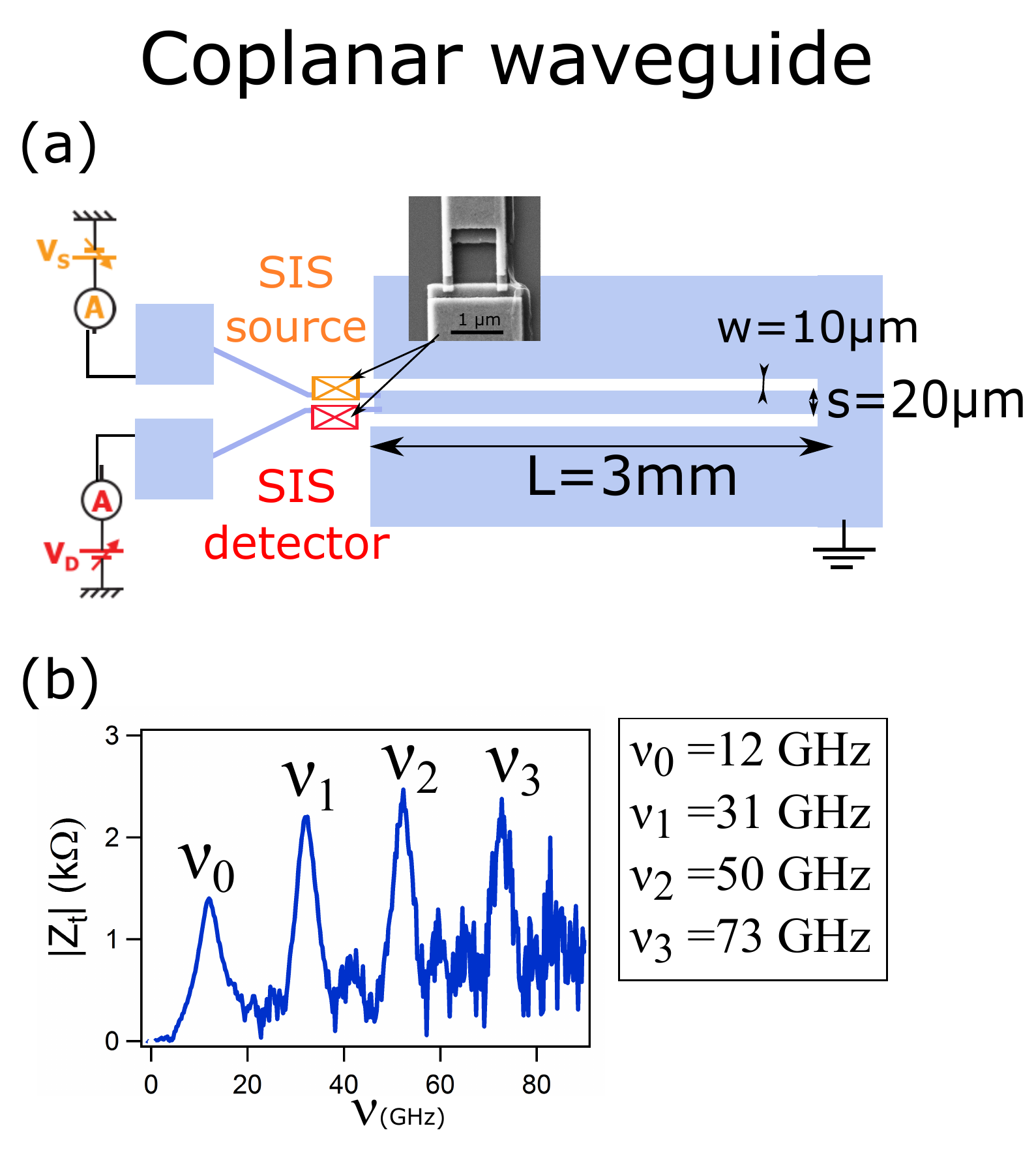}
    \end{center}
    \caption{(a) Design of the coplanar waveguide resonator used in this experiment, showing the coupling between the source and the detector. (b) Measured transimpedance $Z_t(\nu)$, such that the relation between the current fluctuations of the source and the voltage fluctuations across the detector is: $\delta V=|Z_t(\nu)| \delta I$.}
    \label{design_resonators}
    \end{figure}

The usual way for characterizing a resonator, $i.e.$ for determining its resonance frequencies and quality factors, is to measure the frequency dependent reflection coefficient with high frequency electronics. 
But here, the sample is designed to be addressed by DC measurements, AC signal being confined on-chip. The best way to characterize the resonator (and the detector) is thus to use an on-chip AC source. A very convenient one is given by the AC Josephson effect of a Josephson junction: when biased by the voltage $V_s$, there is a AC current $I(t)=I_C\sin(\frac{2eV_s}{\hbar}t)$ in the junction. The associated current spectral density is $S_I(\nu,V_s)=\frac{I_c^2}{4}\left(  \delta\left(\nu-\frac{2eV_s}{h}\right) +\delta \left(\nu+\frac{2eV_s}{h}\right) \right)$. We assume here a quasi-monochromatic Josephson emission. This gives the emission contribution to the photo-assisted tunnelling current, $Z_t(\nu)$ being the transimpedance defined such that the relation between the current fluctuations $\delta I$ of the source and the voltage fluctuations $\delta V$ across the detector is $\delta V=|Z_t(\nu)| \delta I$ :
\begin{equation}
I_{PAT}(V_d,V_s)=\left( \frac{1}{2V_s}\right)^2 \frac{I_c^2}{4}\left|Z_t\left(\frac{2eV_s}{h}\right)\right|^2 I_{qp}^0(V_d+2V_s)
\end{equation}
Thanks to the estimation of the critical current by the Ambegaokar-Baratoff formula \cite{Ambegaokar1963} and knowing the $I(V)$ characteristic in absence of environment $I_{qp}^0$, the measurement of $I_{PAT}$ at a fixed $eV_d>2\Delta-h\nu_0$ gives access to $|Z_t(\nu)|$. The measurement is presented on fig. \ref{design_resonators}b. In case of the Josephson emission with a finite bandwidth, the resonance peak seen in the PAT current results from the convolution of the transimpedance and the finite bandwidth emission.

\subsection{Amplitude of the Josephson emission peak}
\label{subsec:AC_extract}
\begin{figure}[htpb]
\begin{center}
   \includegraphics[width=8.6cm]{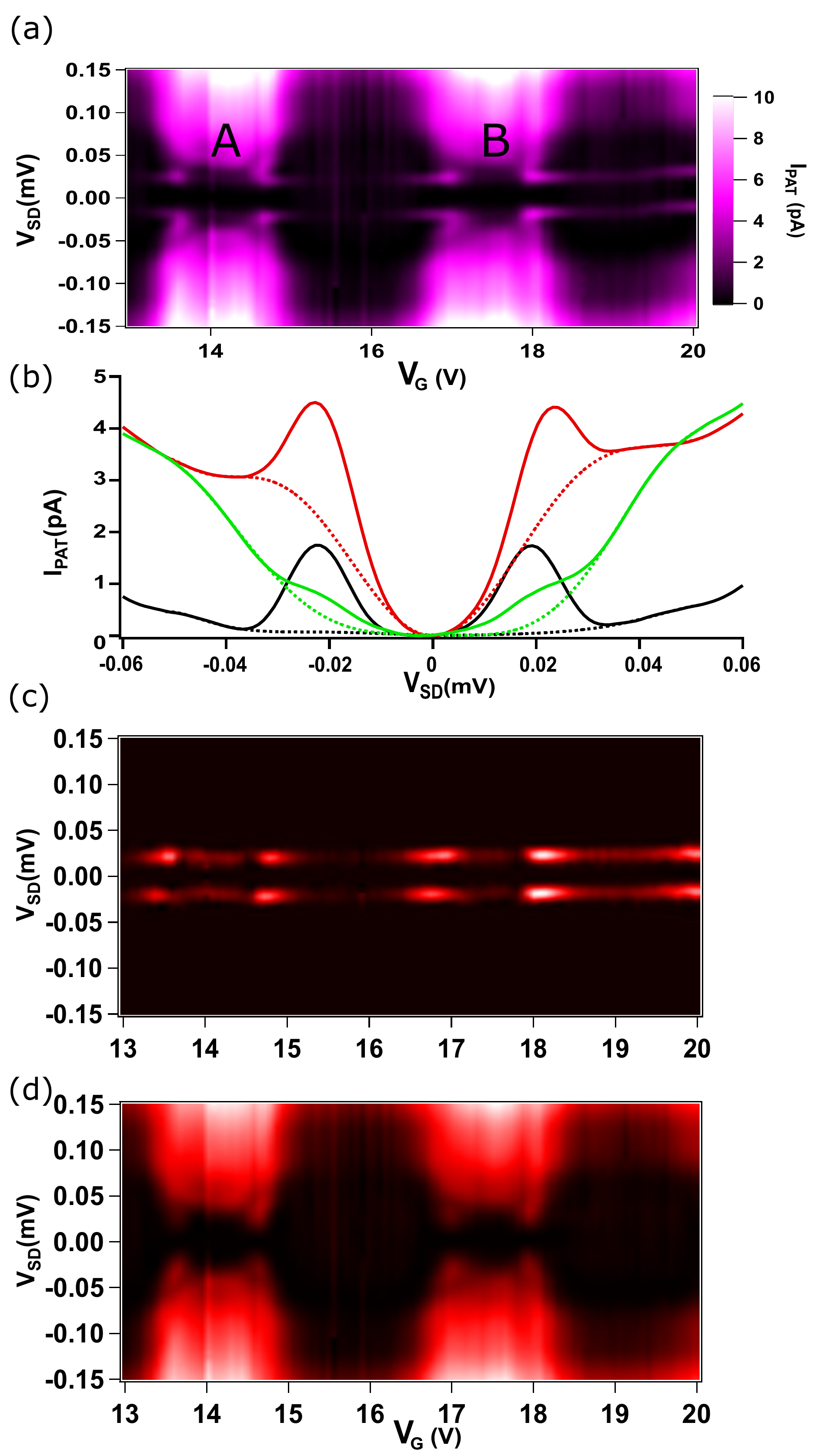}
    \end{center}
    \caption{Extraction procedure of the AC Josephson emission (a) PAT current measured through the detector as a function of the bias voltage $V_{SD}$ and the gate voltage $V_G$ of the CNT quantum dot. (b) The latter signal is separated in a base line and a peak at the Josephson frequency. The base line is obtained by fitting the data away from the Josephson peak with a polynomial. (c) Extracted PAT current related to the AC Josephson effect. It corresponds to the PAT current measured with the base line subtracted. (d) Extracted PAT current corresponding to the MAR process. It correspond to the base line obtained by the procedure described in b.}
    \label{SM_fig5}
\end{figure}

The Josephson emission is extracted from the amplitude of the PAT current through the detector as a function of the gate $V_G$ and bias voltage $V_{SD}$ of the CNT quantum dot. The emission of the carbon nanotube junction has two contributions. The first one is the AC Josephson effect of the CNT junction, at the Josephson frequency given by $h\nu=2eV_{SD}$, and depending on the anharmonicity of the current-phase relation some harmonics. The second contribution is the shot-noise associated to MAR processes and quasiparticle tunneling. In the PAT response, we did not detect any signature of harmonics in the AC Josephson effect. Consequently we separate the two processes by attributing the peak at the Josephson frequency to the AC Josephson effect and the remaining baseline to the shot-noise. This base line is calculated by fitting the data away from the Josephson peak with a polynomial (Fig. \ref{SM_fig5}). This allows to separate the contribution of the AC Josephson effect (Fig. \ref{SM_fig5}c) from the shot-noise (Fig. \ref{SM_fig5}d).

\section{Data for other samples}

\subsection{Data for the Pd/Nb/Al sample}
We show in this part the data on the CNT sample with Pd/Nb/Al. The contacts of the tube are 400 nm apart and made of Pd(8nm)/Nb(11nm)/Al(50nm) trilayer with an effective gap $\Delta=150 \mu eV$, higher than the one of the Pd/Al sample. The presence of a thin layer of Pd provides good contact on the CNTs, however it reduces the superconducting gap compared to that of Al or Nb. For the Pd/Al/Nb contact one has to apply a magnetic field of more than 1T to suppress superconductivity in the contacts. This strongly affect the Kondo resonance and thus prevent a reliable extraction of all the parameters of the quantum dot. The sample is cooled down in a dilution fridge of base temperature 50 mK and measured through low pass filtered lines. The differential conductance is probed with a lock-in technique.

\begin{figure}[htpb]
\begin{center}
   \includegraphics[width=8.6cm]{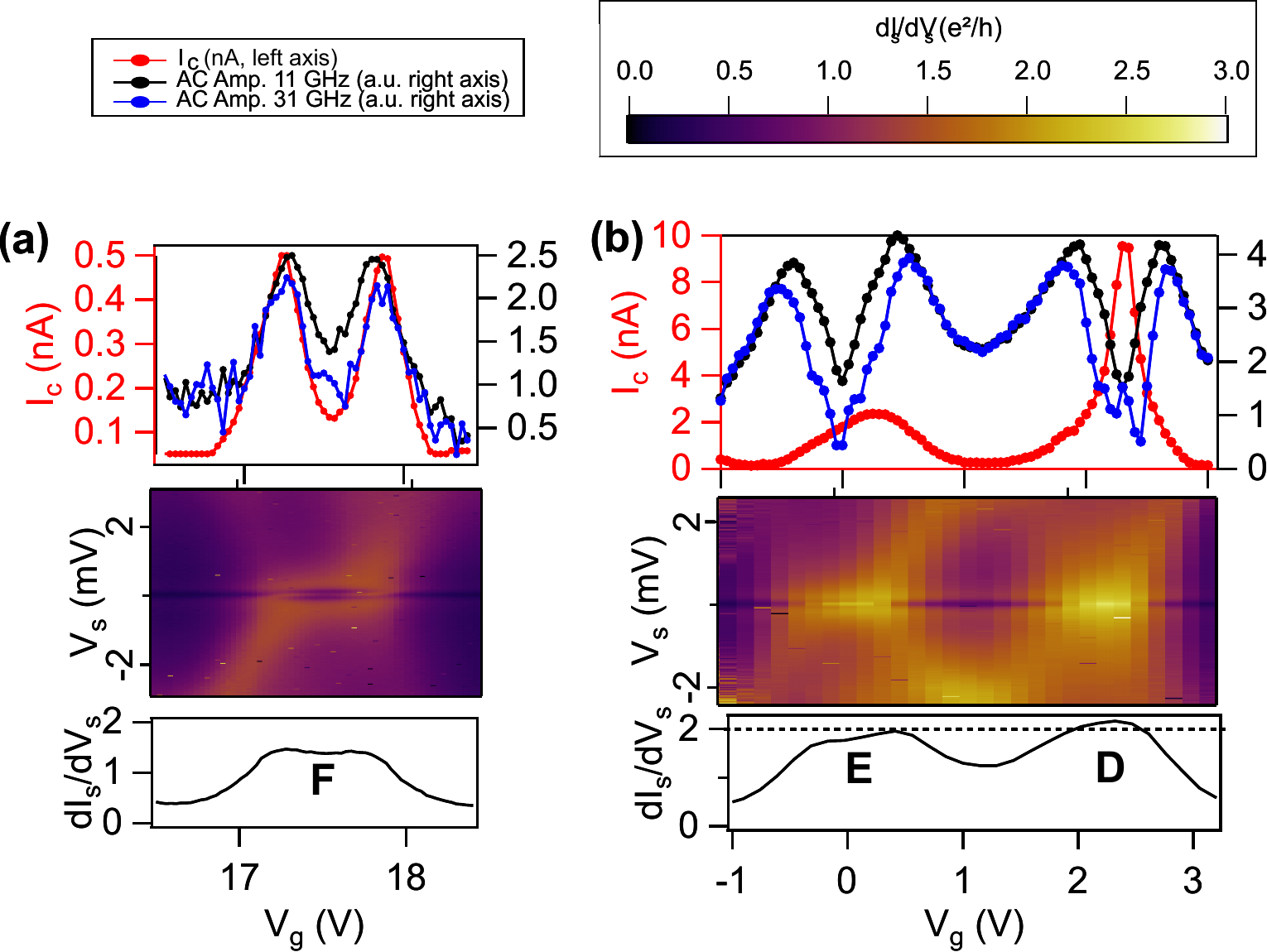}
    \end{center}
    \caption{Comparison between the AC Josephson amplitudes at 11 and 31 GHz and the critical current for the sample with Pd/Nb/Al samples. The region D is the Kondo region described in the main article. Upper panel : the AC amplitudes (11 GHz in black, 31 GHz in blue) are represented in arbitrary unit (right axes), which is the same for the two plots. The critical current is plotted in red, the scale is indicated on the left axes. Middle panel : the differential conductance in presence of a 1T magnetic field is represented as a function of the bias and gate voltages. Lower panel : horizontal cuts of the previous color plots are given for $V_s\approx-0.3\mathrm{~mV}$.}
    \label{SM_fig3}
\end{figure}

On figure \ref{SM_fig3} we show the differential conductance of the CNT quantum dot with a 1T magnetic field applied. The critical current and the AC Josephson emission are compared and exhibit the same qualitative behaviour: a decrease of the Josephson emission for gate regions where the supercurrent is maximal, thanks to the Kondo effect.

\begin{figure}[htpb]
\begin{center}
   \includegraphics[width=8.6cm]{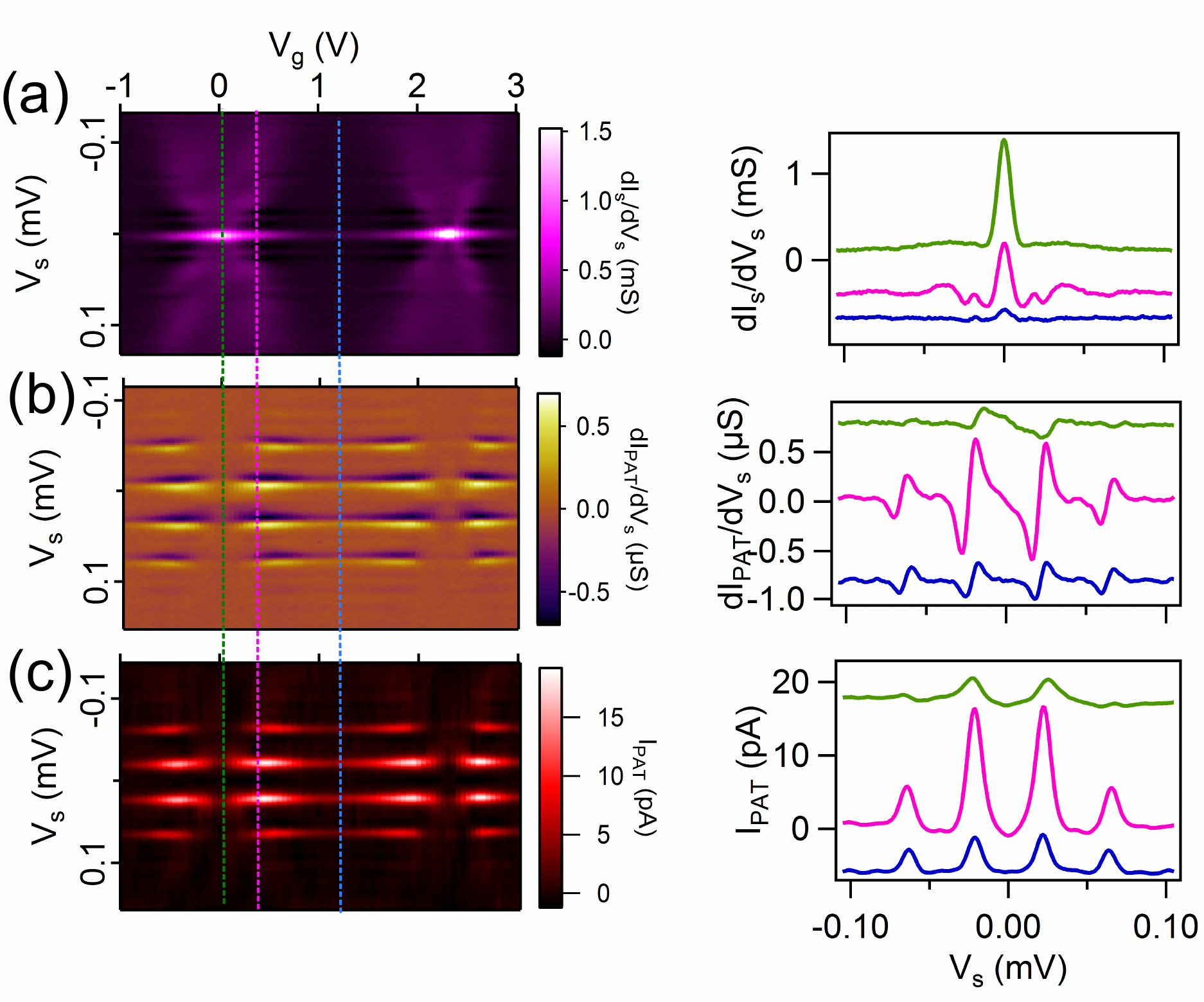}
    \end{center}
    \caption{(a) Differential conductance of the CNT with Pd/Nb/Al contacts as a function of its bias voltage $V_s$ in the superconducting state in the gate voltage region investigated above. (b) Derivative of the photo-assisted tunneling current. (c) Quantity represented on (b) integrated over $V_s$, yielding $I_{PAT}(V_s,V_g)$. Vertical cuts of the three color plots are given on the right at the gate voltages indicated by the dashed color lines. }
    \label{SM_fig3bis}
\end{figure}

The AC Josephson emission is extracted from the value of the PAT current through the SIS detector (Fig. \ref{SM_fig3bis}). The superconducting gap of the trilayer Pd/Al/Nb is higher than the one of the bilayer Pd/Al. This allows the detection of the Josephson emission at the first and third resonance frequency of the coupling circuit, i.e. 11GHz and 31GHz.

\subsection{Data for zone C}

\begin{figure}[htpb]
\begin{center}
   \includegraphics[width=8.6cm]{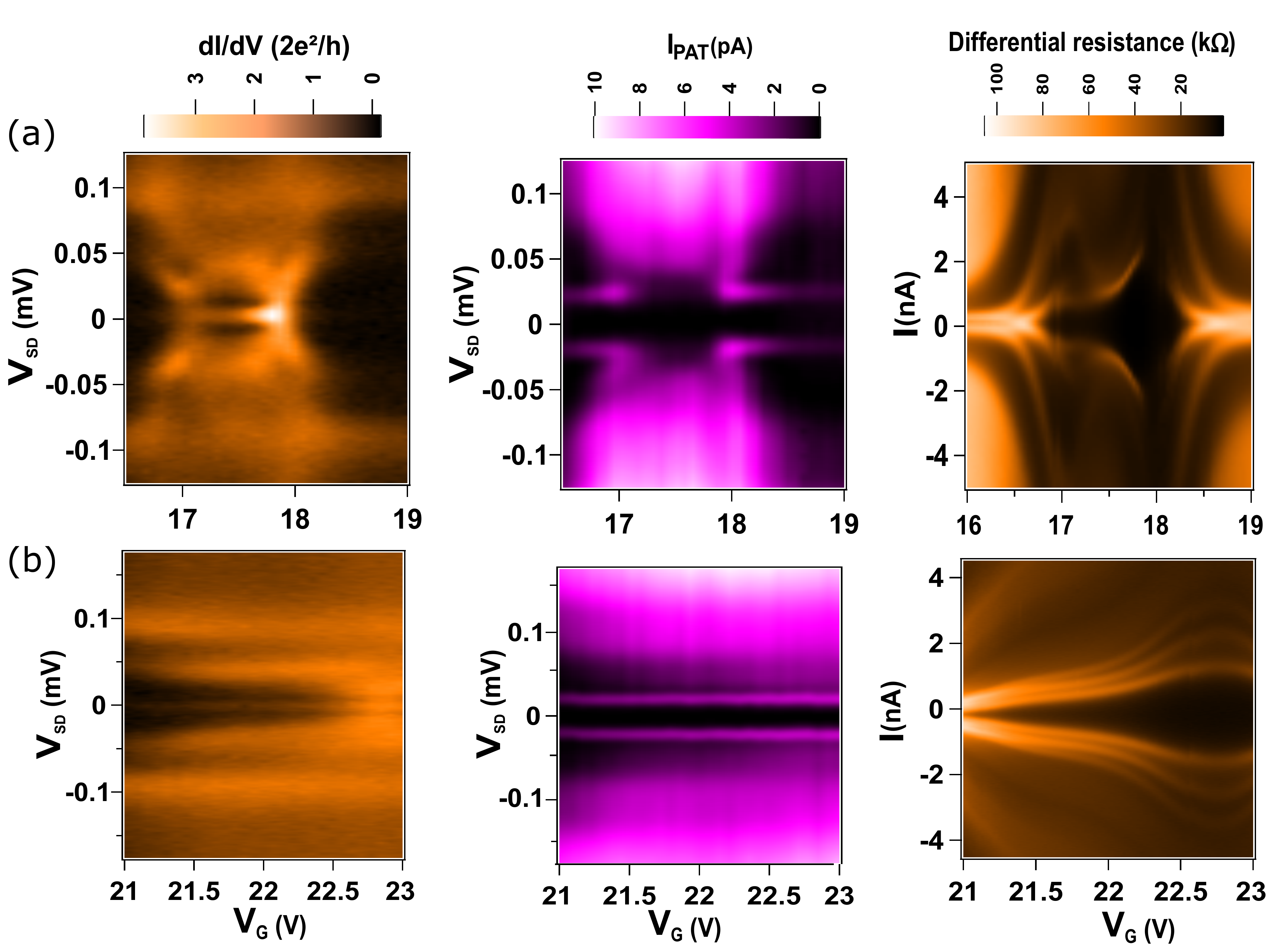}
\end{center}
\caption{Left : Differential conductance $dI/dV$ in the superconducting state for zone B (a) and C (b) as a function of the bias voltage $V_{SD}$ and the gate voltage $V_G$ of the carbon nanotube Josephson junction.  Center : PAT current through the SIS detector. Right : Differential resistance $dV/dI$ as a function of bias current $I$ and gate voltage $V_G$.}
\label{SM_fig4}
\end{figure}

We have also measured the Josephson emission on the sample with Pd/Al contacts in a gate region, called zone C, with a normal state resistance close to the quantum of resistance (Fig. 1b of the main article). This zone does not show any reduction of the Josephson emission, measured via the PAT current through the SIS detector (Fig. \ref{SM_fig4}b). For this region the emission behaves as the supercurrent (fig. 3d of the main article). This points towards the different nature of the process involved for zone C and for Kondo regions A and B. 

\subsection{Comparison of $dI/dV$ and $I_{PAT}$}

\begin{figure}[htpb]
\begin{center}
   \includegraphics[width=8.6cm]{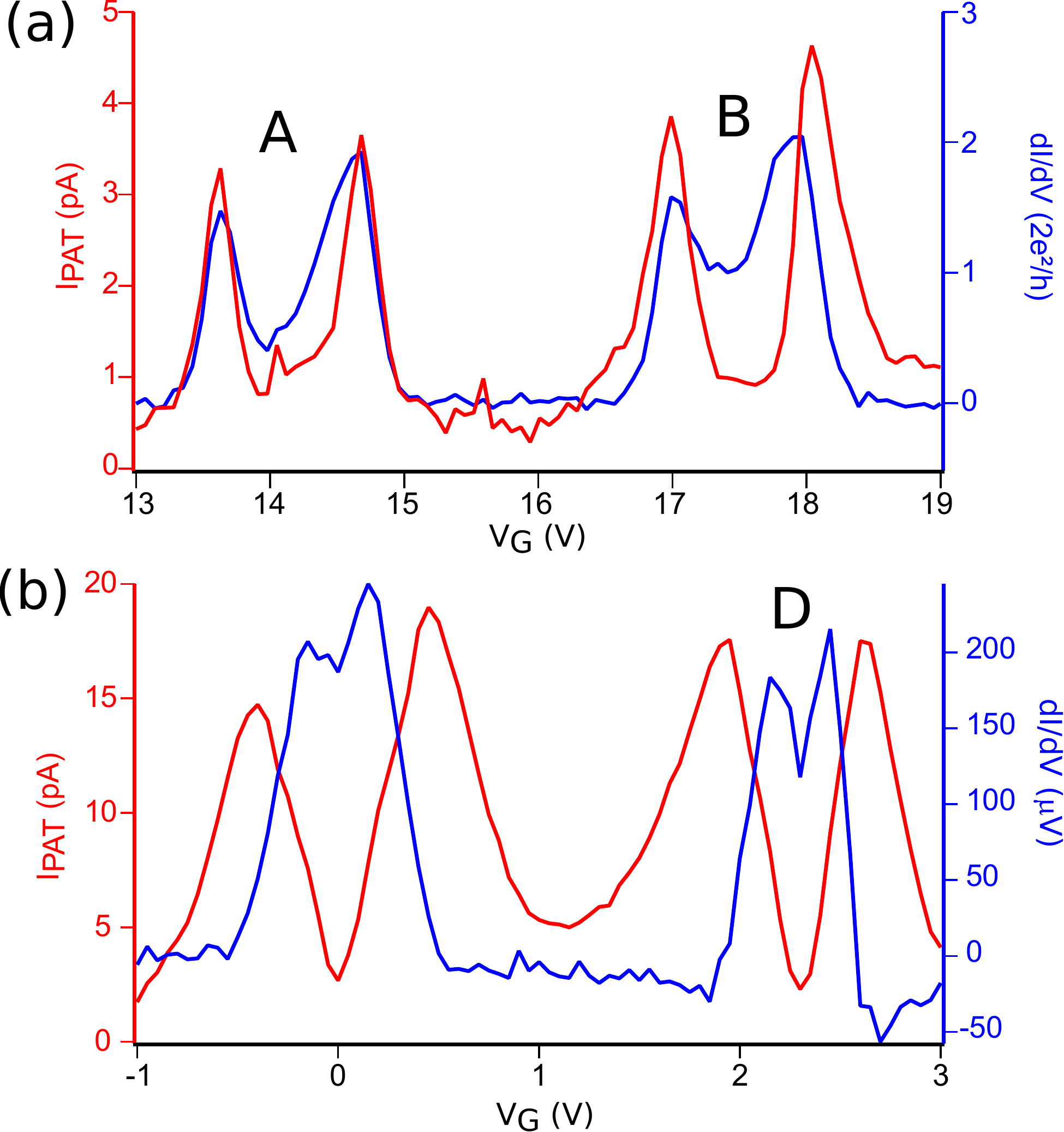}
\end{center}
\caption{Comparison of $dI/dV$ in the superconducting state and $I_{PAT}$ for sample with Pd/Al contacts (upper panel, regions A and B) and for sample with Pd/Nb/Al (lower panel, region D).}
\label{SM_comp_dIdV_Ipat}
\end{figure}

For the sample with Pd/Al contacts, the features in the PAT current associated with the AC Josephson effect and the one associated with multiple Andreev reflection (MAR) are somehow mixed at voltage bias around 20 $\mu$V, due to relatively low value of the superconducting gap ($\Delta = 50 \mu$eV). For region A, $dI/dV$ of the CNT junction in the superconducting state and the PAT current in the detector shows some similarity, this is less true for region B (Figure \ref{SM_comp_dIdV_Ipat}a). The procedure explained in section~\ref{subsec:AC_extract} aims at separating the contributions of the AC Josephson effect and the MAR.

The sample with Pd/Nb/Al contacts has a much higher superconducting gap so that the AC Josephson effect and the MAR processes are well separated as a function of bias voltage (see fig. \ref{SM_fig3bis}a of the SM). Nevertheless this sample exhibits also a reduction if $I_{AC}$ in region where the critical current is maximum due to Kondo effect. On figure \ref{SM_comp_dIdV_Ipat}b, we see that the behaviour of $dI/dV$ and the PAT current in the detector exhibit very different behaviour.  

\section{NRG Calculation of the Andreev spectrum and the supercurrent}

\begin{figure}[tb]
    \begin{center}
    \includegraphics[width=8.6cm]{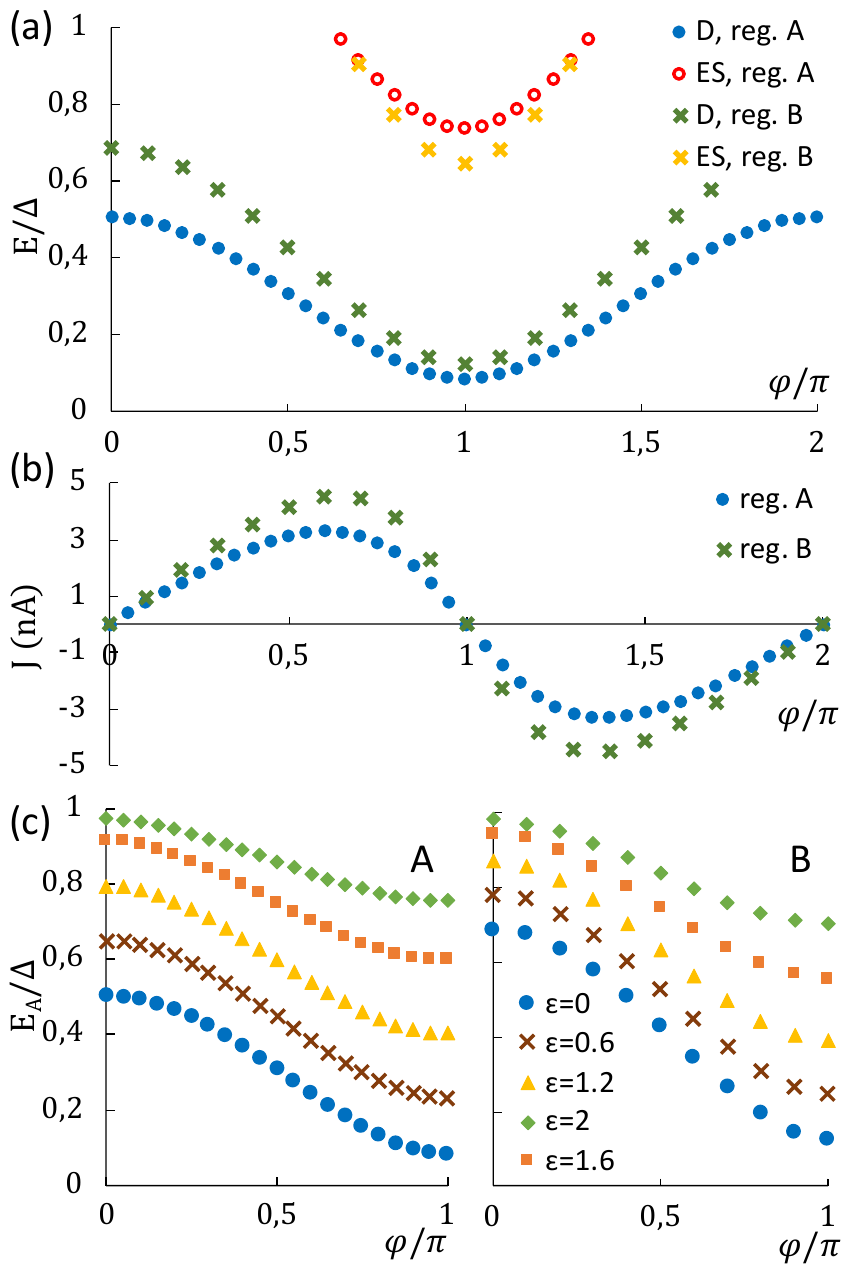}
    \end{center}
    \caption{NRG results. (a) The many-body spectra for Kondo regions A and B and $\varepsilon = 0$. The ground state is a singlet and its energy is set to zero in NRG.  The first excited state is a spin doublet (marked D in the legend) and the second excited state is a singlet (marked ES). This spectrum corresponds to a junction in the 0-phase with one pair of Andreev bound states. (b) The supercurrent of the ground state for Kondo ridge A (blue) and B (green). (c) Andreev bound state energy for different values of the level energy $\varepsilon$ for regions A and B.}
    \label{NRG_results}
    \end{figure}

Numerical renormalization group (NRG) calculations in this work have been performed using the nrgljubljana software \cite{Zitko2009,Zitko2014} and the single-impurity Anderson model. Using the parameters determined in the normal state (table 1 of the main article) for regions A and B, the $\varphi$ and $\varepsilon$ dependent spectra of many-body states and the current-phase relations have been calculated.  Fig. \ref{NRG_results}a is showing the spectrum of excited sub-gap many-body states at half-filling ($\varepsilon=0$), with the ground-state energy equated to zero. The ground state is always a singlet, confirming both regions A and B stay in the 0-phase in the entire range of $\varphi$. The first excited state is the spin doublet, and its energy difference from the ground state corresponds to Andreev bound state energy $E_A$. The second excited state is again a singlet, which is not linked to the ground state by single-particle processes, hence doesn't produce a pair of ABS. Fig. \ref{NRG_results}b shows the corresponding current phase relations, which are slightly non-sinusoidal, with the critical current in the nanoampere range. The $\varepsilon$ dependent spectra for $\varphi=0$ and $\varphi=\pi$ are shown in Fig \ref{Pcontinuum}a-b. With increasing distance from the center of the Coulomb diamond the energy of the excited states increases. 

%To get a quantitative understanding of the behaviour of the quantum dot in region A and B we have performed numerical renormalization group (NRG) calculation \cite{Zitko2009,Zitko2014} of the energy spectrum and supercurrent (Fig. \ref{NRG_results}) using the parameters determined in the normal state (table 1 of the main article). It confirms that the ground state of the system is always the singlet state. This leads to a supercurrent in the nanoampere range, consistent with the experiment, with the phase behaviour of a "0-junction". The current-phase relation is slightly non-sinusoidal. 

\section{Comparison of quantum dot and quantum channel Josephson junctions}

Quantum dots in the Kondo regime have been sometimes treated like
a single quantum channel (or quantum point contact) \cite{Yeyati-03,Vecino-04}.
The idea is that Coulomb interaction $U$ causes a renormalization
of parameters (ABS energies, transmission), but doesn't
produce qualitative differences. As long as the junction remains in
the zero phase in the entire range of the superconducting phase difference
$\varphi\in(0,\,2\pi)$, the structure of many-body levels is indeed
similar, in both cases consisting of a singlet ground state, an excited spin doublet
and an excited spin singlet. There are however important differences. 

\begin{figure}[tb]
\includegraphics[width=8.6cm]{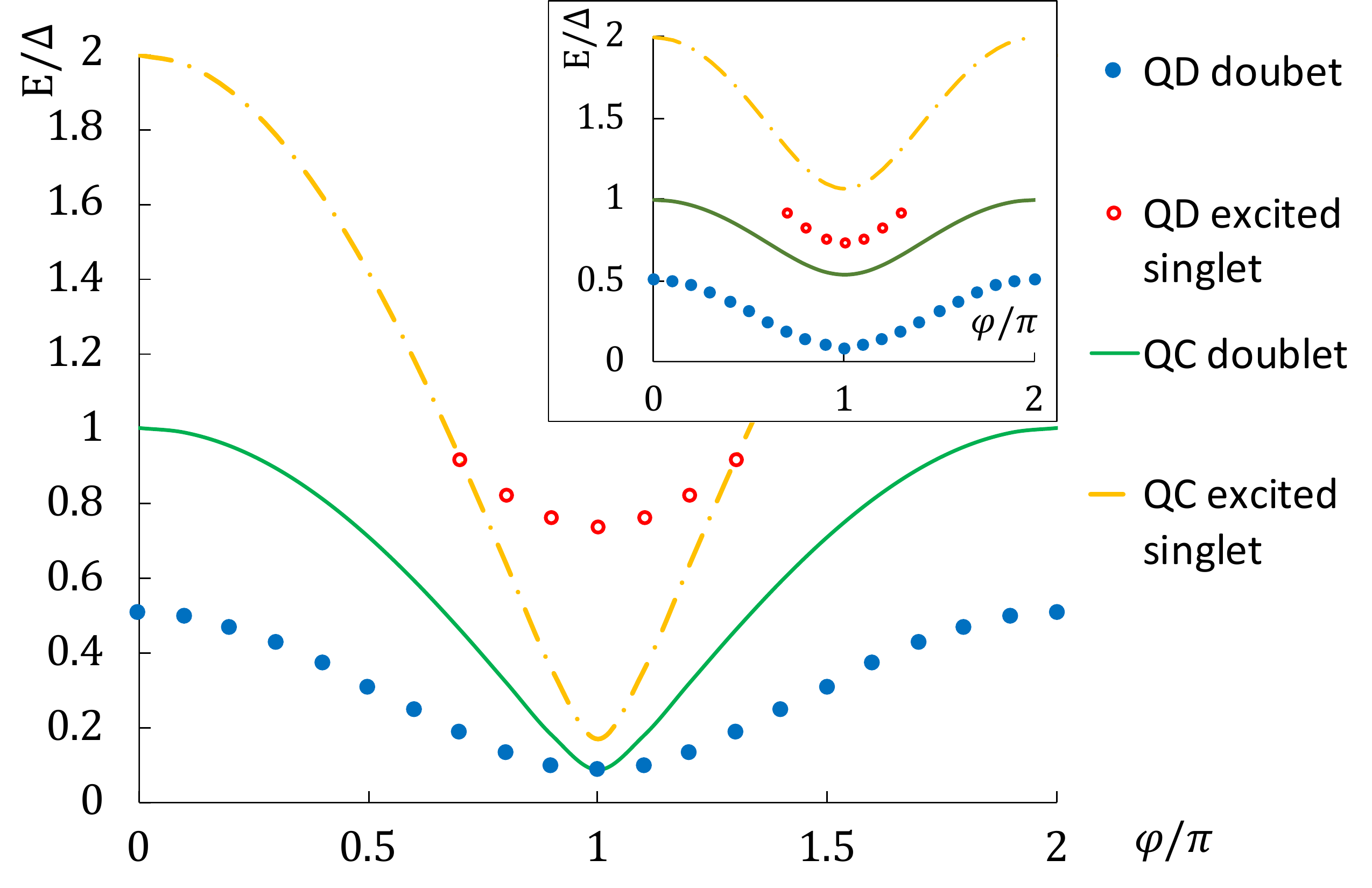} 
\caption{Comparison of the many body spectrum for Kondo ridge A at the particle-hole symmetry point and the spectrum of a quantum channel with the same energy of the Andreev bound state at $\pi$, as a function of the phase difference $\varphi$. The dots represent NRG data for the energy difference between the spin doublet and the ground state, corresponding to the ABS energy (blue) and the difference between the excited singlet and the ground state (red), the solid lines represent the same quantities for the quantum point contact. The ABS for the quantum dot detach from the continuum (starting at $\Delta$) at $\varphi=0$, and the energy of the excited singlet for the QD is much higher then that of a QC at $\varphi=\pi$. Note that, due to technical reason, the NRG calculation does not give the energy of the singlet state for values higher than the gap $\Delta$. Inset: Comparison of the many body spectrum for Kondo ridge A at the particle-hole symmetry point and the spectrum of a quantum channel with the same conductance in the normal state.}
\label{QCvsQD} 
\end{figure}

\begin{figure}[tb]
\begin{centering}
\includegraphics[width=8.6cm, angle = 0]{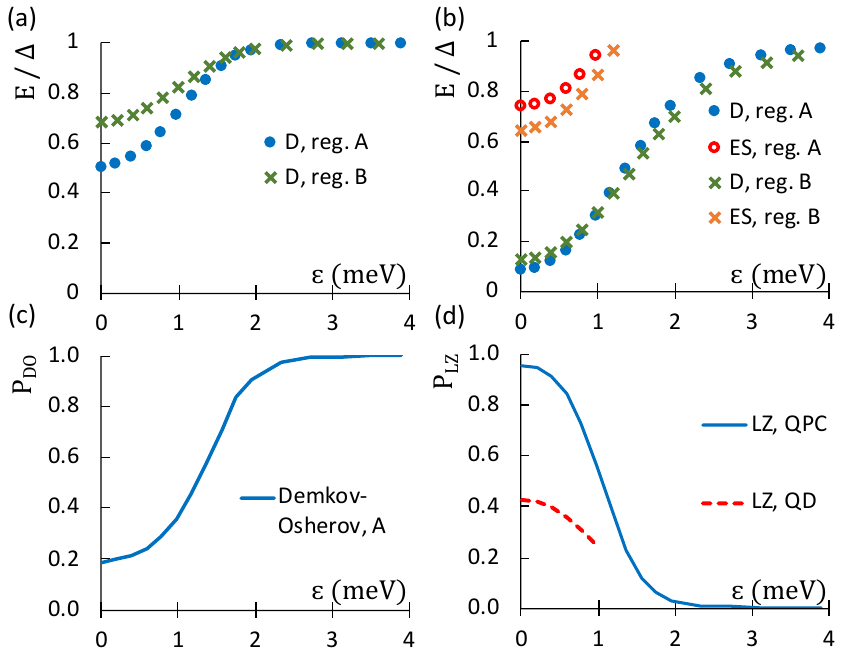} 
\par\end{centering}
\caption{(a) NRG calculation of the many-body spectrum (ABS energy) at $\varphi=0$ as a function
of the energy level of the QD $\epsilon$ for for Kondo regions A and B. This
measures the detachment of the ABS from the continuum of excitation.
(b) Same quantity at $\varphi=\pi$. (c) Probability for a QP present
in the quantum dot to escape after tunnelling into the continuum due
to Demkov-Osherov tunnelling. This curve is calculated at a voltage
$eV$=$\Delta/2$ and use the result derived in ref.\cite{Badiane2013,Houzet2013}.
(d) Landau-Zener probability for a QPC model and a QD, calculated based 
on the many-body spectrum of Kondo region A (cf. (b)). }
\label{Pcontinuum} 
\end{figure}

For a short quantum channel the ABS energy (the difference in energy
between the doublet excited state and the ground state) is $E_{A}=\Delta\sqrt{1-\tau\sin^{2}\left(\varphi/2\right)}$
(with $\tau$ the transmission of the junction in the normal state),
and the difference between the energy of the excited singlet and the
singlet ground state is $2E_{A}$. With interaction, none of this
is true any longer: the ABS detaches from the continuum at $\varphi=0$,
the energy of ABS at $\varphi=\pi$ no longer corresponds to the normal
state transmission, and the energy difference between the excited
singlet and the ground state is significantly higher than $2E_{A}$.
A comparison of the ($\varphi$ dependent) many body spectrum of a
quantum point contact vs. NRG data for the interacting quantum dot
 is shown in Fig. \ref{QCvsQD}. We choose to equate the ABS energy at $\varphi=\pi$, corresponding to a (renormalized)
normal state transmission of the quantum channel, $\tau=1-\left(E_{A}(\varphi=\pi) / \Delta \right)^{2}$.
NRG calculations also allow to evaluate the evolution of the detachment
of the ABS from the continuum (Fig \ref{Pcontinuum}a) and the value
of the ABS at $\varphi=\pi$ (Fig \ref{Pcontinuum}b) with changing
level energy $\text{\ensuremath{\varepsilon}}$. In the following
subsections, we give two quantum-channel-based interpretations that
seem plausible until a closer look. 

\subsection{Landau-Zener tunnelling}

In a quantum channel the variation of $I_{C}^{AC}(V_{SD})$ at low
bias voltage has been attributed to Landau-Zener tunnelling \cite{Avernin-95}.
While this is sometimes pictured as particles tunnelling from one ABS
to another, on the many body level, it is a transition from the ground
state to the excited singlet state (the jump between the ground state
and the spin doublet is forbidden by parity). The probability for
this transition to occur is given by \cite{Mullen1988} :
\begin{equation}
P_{LZ}=\exp\left[-\text{\ensuremath{\pi}}\frac{(\delta E/2\Delta)^{2}\Delta}{eV_{SD}}\right].
\end{equation}
We denote $\delta E$ the energy difference between the states involved
in LZ tuneling. For the quantum chanel $\delta E=2E_{A}(\varphi=\pi)$,
and the quantity $(\delta E/2\Delta)^{2}=R=1-\tau$ is the reflectivity
of the junction. $V_{SD}$ is the applied voltage, which determines
the phase evolution through the Josephson relation $d\varphi/dt=2eV_{SD}/\hbar$,
and in the experiment $eV_{SD}\cong\Delta/2$. With the shape of the
ABS pictured in Fig. \ref{QCvsQD} and Fig. \ref{Pcontinuum}b, the
Landau-Zener probability for a quantum channel is close to one at
half-filling and drops to zero far away from it, see Fig. \ref{Pcontinuum}d. Moreover, although
the ABS detach from continuum at $\varphi=0$, the emptying of ABS
states at $\varphi=0$ still happens through Demkov-Osherov tunnelling
processes. The probability $P_{DO}$ of tunnelling between ABS and the
continuum, based on Ref. \cite{Badiane2013,Houzet2013}, is pictured on Fig. \ref{Pcontinuum}c.
Recent work \cite{Lamic2020} has investigated a quantum dot
junction with ABS detached from continuum but still keeping $\delta E=2E_{A}(\varphi=\pi)$
as a Markov chain and found that the ratio between $P_{DO}$
and $P_{LZ}$ is significant for the occupation of the states in the junction.
In our case $P_{DO}<P_{LZ}$ up to $\varepsilon\cong1.1$ meV, which
is where the biggest changes in the experimentally measured current
occur. These observations make it very compelling to call Landau-Zener
responsible for the measured drop in $I_{C}^{AC}$. However, for the
quantum dot, one must take $\delta E$ to be the energy of the excited
singlet ($\delta E(\varepsilon=0,\varphi=\pi)=0.74\Delta$ for the
Kondo ridge A at half-filling), leading to a reduced transition probability
$P_{LZ}^{QD}=0.43$. Moreover, this value changes slowly when
one goes away from the particle-hole symmetry point (Fig. \ref{Pcontinuum}d). 
Hence at $\epsilon=1$meV, where we see in the experiment that the dynamical supercurrent increases,
this Landau-Zener probability is still 0.24 - which makes Landau-Zener
tunnelling in and of itself unsuitable to explain the observed data. 

Instead of equating ABS energy at $\text{\ensuremath{\varphi}=\ensuremath{\pi}}$
and thus the renormalized transmission, one could alternatively consider
a quantum channel with the same normal state transmission as experimentally
measured for the quantum dot (see inset of Fig. \ref{QCvsQD}). With a conductance of $0.71\times2e^{2}/h$
this leads to $\delta E=1.07\Delta$ and a small transition probability
$P_{LZ}=0.16$, thus making the Landau-Zener probability an even less
likely explanation. 

\subsection{A renormalized-quantum-point-contact-based interpretation}

While a tunneling approach gives a simple picture, it is dependent
on the validity of the adiabatic theory (assuming that for small enough
bias voltage the equilibrium levels are still good quantum levels),
which, strictly speaking, is only true for very small voltages. A
full microscopic description is called for, but unavailable for the
interacting quantum dot, therefore, once again we revert to treating
the system like a quantum channel with renormalized transmission.

\begin{figure}[tb]
\includegraphics[width=8.6cm]{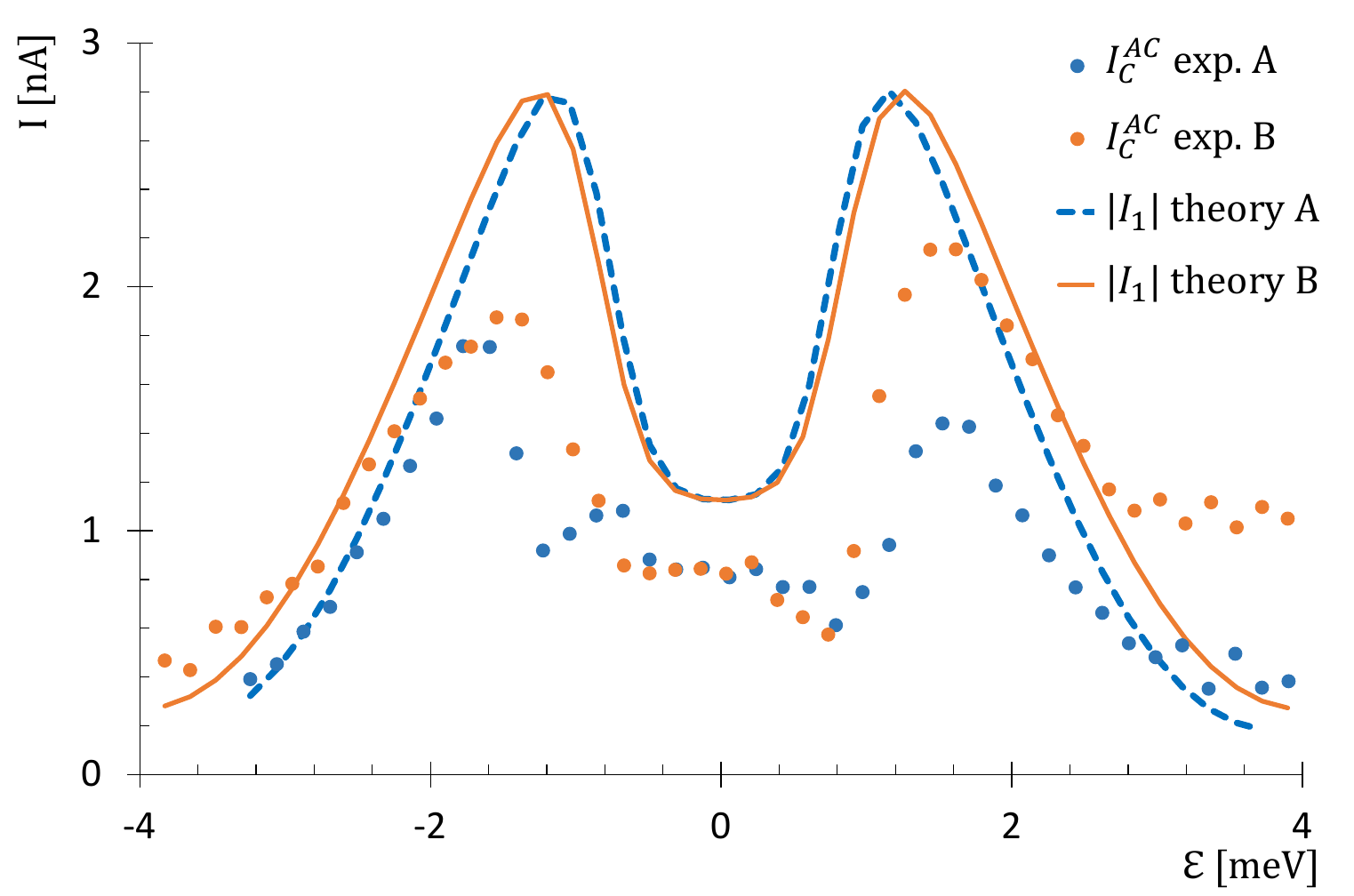} 
\caption{Gate dependence of the first Fourier component $\left|I_{1}\right|$
of the AC Josephson current. Bullets represent the experiment, lines
the theoretical prediction for a single quantum channel with a renormalized
transmission. Transmission of the quantum dot has been evaluated from
the energy of Andreev bound states at $\varphi=\pi$, obtained by
the NRG. The dots represent experimental data for Kondo ridges A and
B. The values of $\left|I_{1}\right|$ are based on results of Refs.
\cite{Avernin-95,Cuevas-96}.}
\label{fig:Comparison} 
\end{figure}

The full transport theory for a short superconducting quantum channel,
has been published in the nineties by Averin and Bardas \cite{Avernin-95}
and Cuevas, Mart\'in-Rodero and Levy Yeyati \cite{Cuevas-96}. These
studies feature results for the real and imaginary part of the first
Fourier component $I_{1}$ of the AC current for several values of
transmission. We use their results (read off graphically) for applied
bias voltage $V=\Delta/2e$ to construct the $\left|I_{1}(\text{\ensuremath{\tau}})\right|$
dependence of the AC current on transmission. The renormalized transmission
of our setup is given by the energy of the Andreev bound states at
$\varphi=\pi$, namely $\tau\left(\epsilon\right)=1-\left(E_{A}(\epsilon,\,\varphi=\pi)/\Delta\right)^{2}$.
We obtain equilibrium values of $E_{A}(\epsilon,\,\varphi=\pi)$
from the NRG, cf. Fig. \ref{Pcontinuum}b. Results for both Kondo
ridges A and B are similar and given in Fig. \ref{fig:Comparison},
showing a nice semi-quantitative agreement between the renormalized-quantum-point-contact
based prediction and measured experimental data. However, this analysis
does not take into account the differences in the many-body spectrum
of the quantum channel vs. quantum dot (illustrated in equilibrium
by the detachment of ABS from the continuum and the raised energy
of the excited singlet), which raises significant doubts about its
accuracy. 

\section{Evaluation of the quasiparticle dynamics in the QD junction}

\begin{figure}[tb]
    \begin{center}
    \includegraphics[width=8.6cm]{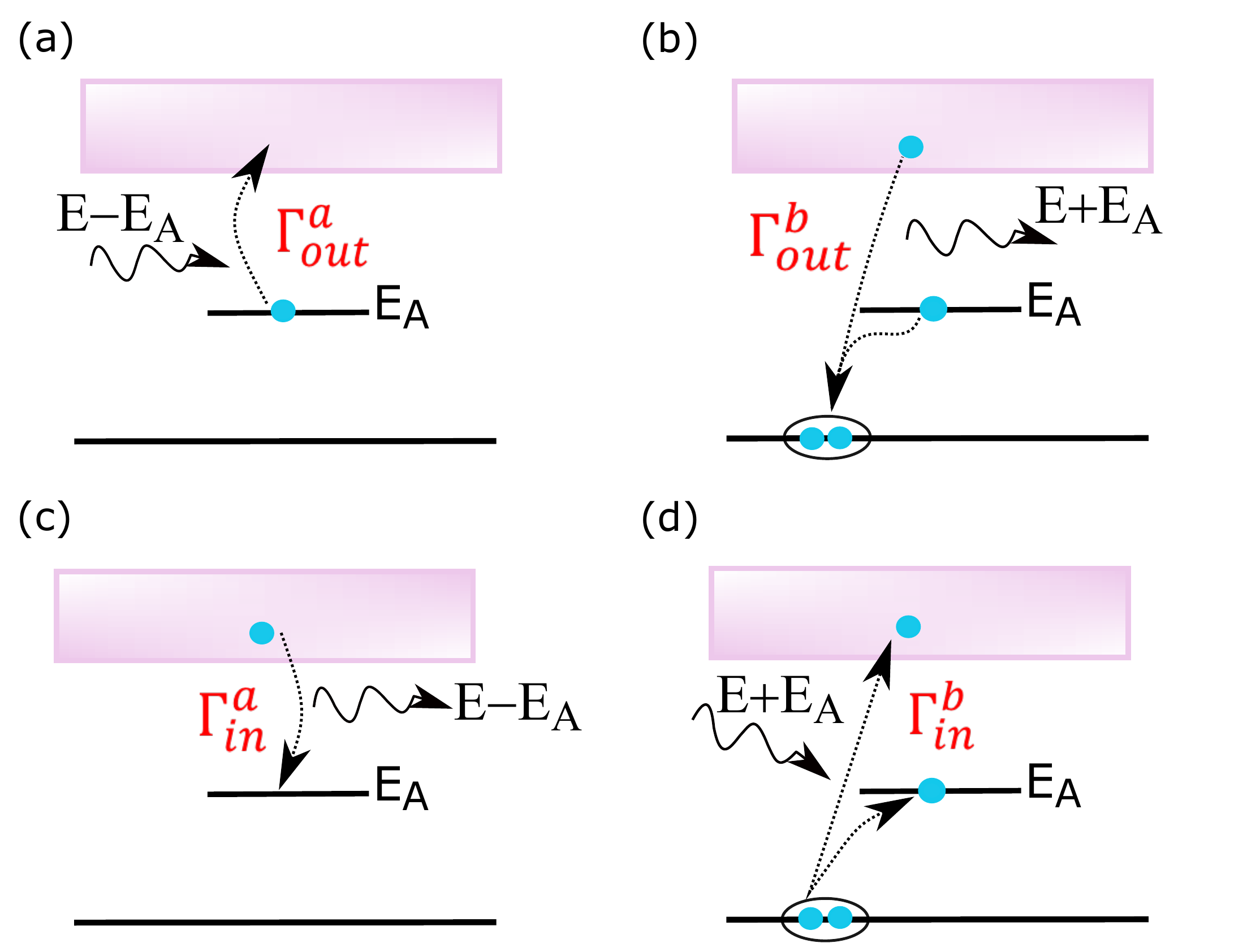}
    \end{center}
    \caption{Sketch of the processes corresponding to the different rates controlling the dynamics of the QP in the QD.}
    \label{fig:poisoning}
\end{figure}

We discuss here the quasiparticle (QP) dynamics, which may lead to
the occupation of the doublet state in the QD Josephson junction.
We evaluate the different rates controlling the injection and escape
of QP in the dot \cite{Olivares2014} due to the effect of the electromagnetic
environment of the junction. This environment is, in the present experiment,
constituted by the resonant coupling circuit of impedance $Z_{env}$.
Following Ref. \cite{Olivares2014} we consider two processes and
their time-reversed versions (figure \ref{fig:poisoning}): a) the escape of a QP from the Andreev
level to the continuum after absorbing a photon from the environment,
and b) the recombination of a QP at the Andreev level with a QP from
the continuum into a Cooper pair while emitting excess energy. The
probability to absorb energy from the environment is described by
$P(E)=D(E)f_{BE}(E,T_{env})$, where $D(E)$ is the density of states
in the environment and $f_{BE}(E,\,T_{env})$ the Bose-Einstein distribution
at energy $E$ and temperature $T_{env}$. The density of states is
given by $D(E)=Re(Z_{env}(E)/E)/R_{Q}$, with $R_{Q}=h/4e^{2}$. Starting
with Fermi golden rule, Ref. \cite{Olivares2014} arrives at the following
rates:
\begin{itemize}
\item The rate for a QP on the Andreev level with energy $E_{A}$ to escape
into the continuum at the energy $E$ after absorbing a photon with
energy $E-E_{A}$ from the environment:
\begin{align*}
\Gamma_{out}^{a} & =\frac{8\Delta}{h}\int_{\Delta}^{+\infty}dED(E-E_{A})g(E,E_{A})\\
 & \ \times f_{BE}(E-E_{A},\,T_{env})(1-f_{FD}(E,\,T_{qp}))
\end{align*}
Here $f_{FD}(E)$ is the Fermi-Dirac function, describing the QP in
the continuum at a temperature $T_{qp}$ and the function $g(E,E_{A})$
is related to the matrix element of the current operator and is approximated
by $g(E,E_{A})=\sqrt{(E^{2}-\Delta^{2})(\Delta^{2}-E_{A}^{2})}/[\Delta(E-E_{A})]$
(which is precise in the $\tau\rightarrow1$ limit). 

\item The rate recombination of a QP at the Andreev level $E_{A}$ with
a QP from the continuum into a Cooper pair while emitting excess energy
$E+E_{A}$ into the environment:
\begin{align*}
\Gamma_{out}^{b} & =\frac{8\Delta}{h}\int_{\Delta}^{+\infty}dED(E+E_{A})g(E,-E_{A})\\
 & \ \times(1+f_{BE}(E+E_{A},\,T_{env}))f_{FD}(E,\,T_{qp})
\end{align*}

\item The rate for a particle of energy $E$ to enter the QD and occupy
the Andreev level, after emitting energy $E-E_{A}$ :
\begin{align*}
\Gamma_{in}^{a} & =\frac{8\Delta}{h}\int_{\Delta}^{+\infty}dED(E-E_{A})g(E,E_{A})\\
 & \ \times(1+f_{BE}(E-E_{A},\,T_{env}))f_{FD}(E,\,T_{qp})
\end{align*}

\item The rate for breaking a Cooper pair into one QP occupying the Andreev
level and another one in the continuum at energy E, after absorbing
the energy $E+E_{A}$ from the environment:
\begin{align*}
\Gamma_{in}^{b} & =\frac{8\Delta}{h}\int_{\Delta}^{+\infty}dED(E+E_{A})g(E,-E_{A})\\
 & \ \times f_{BE}(E+E_{A},\,T_{env}))(1-f_{FD}(E,\,T_{qp}))
\end{align*}
\end{itemize}

\begin{figure}[tb]
\begin{centering}
\includegraphics[width=8.6cm]{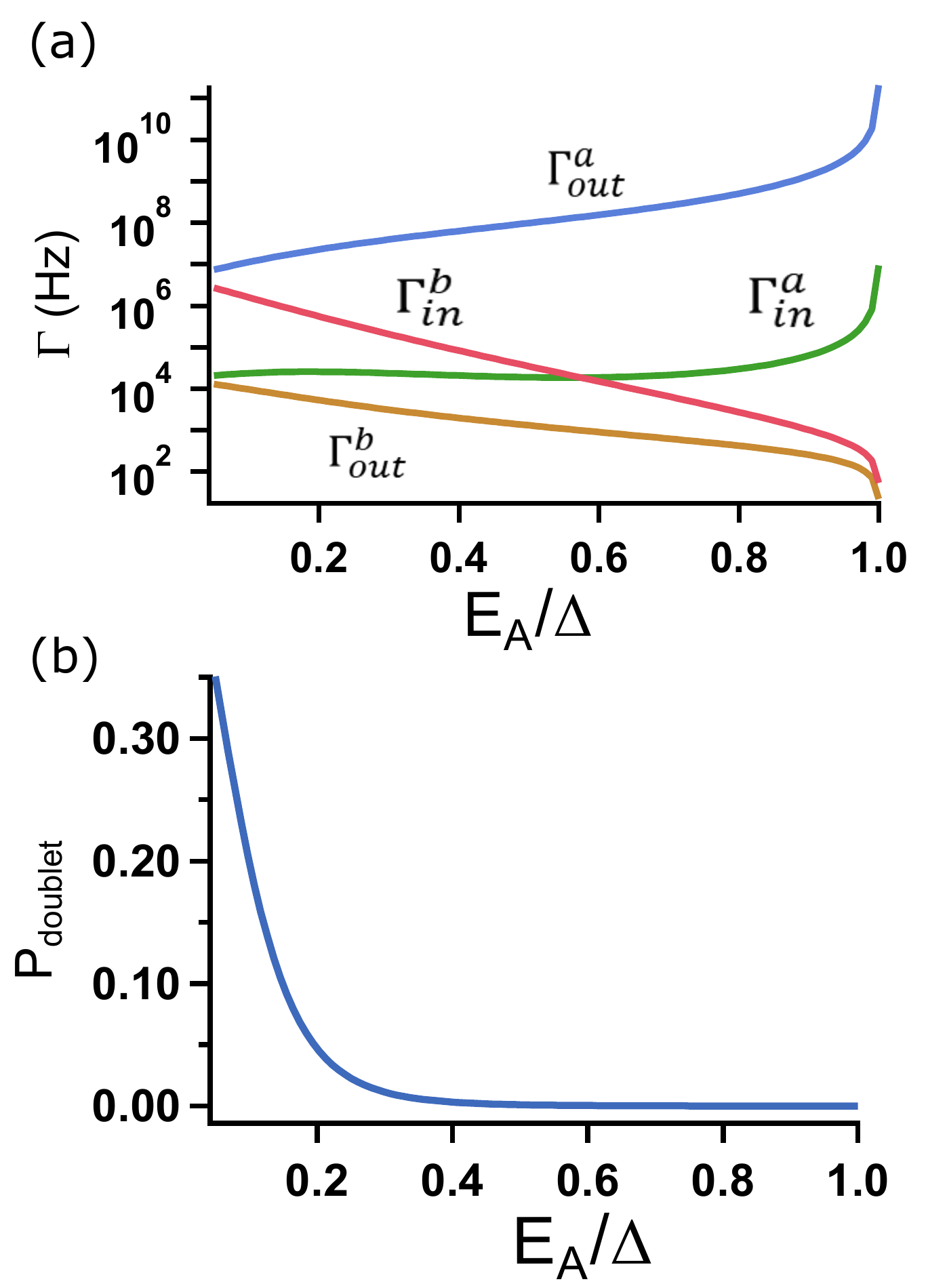} 
\par\end{centering}
\caption{(a) Numerical evaluation of the rate of the QP injection and escape
in the QD junction with $k_{B}T_{env}=0.2\Delta$ and $k_{B}T_{qp}=0.1\Delta$.
(b) Probability for the junction to be in the doublet state as a function
of the energy $E_{A}$ of the Andreev level.}
\label{LZproba} 
\end{figure}

These rates can be evaluated numerically, the results are pictured
in Fig. \ref{LZproba}a, where we have chosen $k_{B}T_{env}=0.2\Delta\approx120$mK
and $k_{B}T_{qp}=0.1\Delta\approx60$mK. The probability to be in
the doublet state is computed as $P_{D}=2\Gamma_{in}/(3\Gamma_{in}+\Gamma_{out})$
with $\Gamma_{in(out)}=\Gamma_{in(out)}^{a}+\Gamma_{in(out)}^{b}$,
and plotted as a function of the position of the Andreev level $E_{A}$
in Fig. \ref{LZproba}b. For an energy of the Andreev level higher
than $0.2\Delta$, the probability is extremely small (below 0.05),
thus explaining why there is no sign of the doublet state in the DC
supercurrent measurement. As stated in the main text, we expect $P_{D}$
to be significantly higher in a voltage-biased situation.